\documentclass[nofootinbib,aps,pra,amsmath,amssymb,amsfonts,superscriptaddress,floatfix]{revtex4} 
\pdfoutput=1
\usepackage{amsmath,amsthm,amsfonts,amssymb}
\usepackage{graphicx}
\usepackage{color}
\usepackage{hyperref}

\newtheorem{theorem}{Theorem}

\newtheorem{lemma}{Lemma}

\newtheorem{definition}{Definition}

\usepackage{algorithm}
\usepackage[capitalize,nameinlink]{cleveref}

\newcommand{\dev}{{\rm dev}}

\newcommand{\supp}{{\rm supp}}
\newcommand{\be}{\begin{equation}}
\newcommand{\ee}{\end{equation}}

\newcommand{\cC}{{\cal C}}
\newcommand{\cD}{{\cal D}}
\newcommand{\cE}{{\cal E}}

\newcommand{\poly}{{\rm poly}}
\newcommand{\polylog}{{\rm polylog}}

\newcommand{\cS}{{\cal S}}
\newcommand{\cF}{{\cal F}}
\newcommand{\cB}{{\cal B}}

\newcommand{\cQ}{{\cal Q}}
\newcommand{\cR}{{\cal R}}
\newcommand{\cX}{{\cal X}}
\newcommand{\cA}{{\cal A}}

\newcommand{\mZ}{{\mathbb Z}}
\newcommand{\Cn}{\operatorname{Cone}}
\begin{document}

\title{On Quantum Weight Reduction}
\author{M. B. Hastings}

\affiliation{Station Q, Microsoft Research, Santa Barbara, CA 93106-6105, USA}
\affiliation{Microsoft Quantum and Microsoft Research, Redmond, WA 98052, USA}
\begin{abstract}
We give a general procedure for weight reducing quantum codes.  This corrects a previous work\cite{owr}, and introduces a new technique that we call ``coning" to effectively induce high weight stabilizers in an LDPC code.
As one application, any LDPC code (with arbitrary $O(1)$ stabilizer weights) may be turned into a code where all stabilizers have
weight at most $5$ at the cost of at most a constant factor increase in number of physical qubits and constant factor reduction in distance.
Also, by applying this technique to a quantum code whose $X$-stabilizers are derived from a classical log-weight random code and whose $Z$-stabilizers have linear weight, we construct an LDPC quantum code with distance $\tilde \Omega(N^{2/3})$ and $\tilde\Omega(N^{2/3})$ logical qubits.
\end{abstract}
\maketitle

An LDPC quantum code is a code (more precisely, a family of codes) in which every stabilizer acts on $O(1)$ qubits and every qubit participates in $O(1)$ stabilizers.  
A natural goal is to construct LDPC quantum codes with linear distance and positive rate; classical codes with such properties are called ``good" and many constructions of them are known.
For quantum LDPC codes, there are several constructions achieving
distance $N^{1/2} \operatorname{polylog}(N)$ based on ideas from topology~\cite{FML02}
and higher-dimensional expanders~\cite{EKZ20,KT20}.  Finally the $N^{1/2} \polylog(N)$ barrier was broken.  A construction based on fiber bundles\cite{hho} achieved distance
$\Omega(N^{3/5}/\polylog(N))$.  Then, another construction gave linear distance up to a logarithm with logarithmically many logical qubits\cite{LD}, allowing
construction of codes with 
distance $N^{1-\alpha/2}$ and $N^{\alpha}$ logical qubits for any $\alpha\in (0,1]$.  Finally,
the fiber bundle construction was derandomized\cite{BE}, and the rate was improved, giving a different construction of codes
with distance $N^{1-\alpha/2}$ and $N^{\alpha}$ logical qubits, at least for $\alpha\in [4/5,1]$.

The present paper takes a different approach, based on weight reduction.  
Weight reduction, as the name implies, is a procedure to reduce the weight of stabilizers in a quantum code (and reduce the number of stabilizers acting on each qubit), perhaps paying the price of some reduction in distance and some increase in the number of physical qubits.
This price is only polynomial in the ``deviation from LDPC" of the original code, suitably defined later.
 
This has several possible applications; we discuss four.
First, many of the constructions of LDPC quantum codes with distance $\omega(N^{1/2})$ have fairly high weight stabilizers: while the weights are $O(1)$, the actual numbers are quite large.  By weight reduction, we can construct new codes with stabilizer weights bounded by some universal quantities which are small $O(1)$ numbers, paying only a constant factor multiplicative cost in rate and number of physical qubits\footnote{However, current constructions based on products and twisted products of classical codes including \cite{hho,LD,BE} as well as \cite{TZ14} can be weight reduced more simply using the classical weight reduction of \cite{hho} which has the further advantage of being a chain homotopy equivalence with bounded Lipschitz constants.  In more detail: for hypergraph product codes which use untwisted products, we can weight reduce both classical codes in the construction.  For codes which are fiber bundles, we can weight reduce the base code, and the fiber used in current constructions is a circle which has low weight.}.
Second, in this paper we give a construction of a code
with distance $N^{1-\alpha/2}$ and $N^\alpha$ logical qubits (up to polylogs) for $\alpha\in[2/3,1]$.   This is based on weight reducing a (fairly simple) good quantum code with constant weight $X$-stabilizers and high weight $Z$-stabilizers.
It is perhaps rather
surprising that such a different construction gives similar scaling as those constructions above, which are instead based on (generalized) products of complexes.  This similarity could perhaps be interpreted as evidence that there is some fundamental limit to distance and rate of quantum LDPC codes and that these codes are reaching this limit; alternatively, it could be interpreted optimistically, that further improvements may be possible.
A third conjectural application is to other constructions of good quantum codes which are almost, but not quite, LDPC.
As an example, consider the homological product
 codes of \cite{bh}; these codes have linear distance and rate and square-root weight stabilizers and are constructed by a homological product of two random codes.  If one could change them into linear distance and rate codes on $N$ qubits in which every stabilizer has weight $O(N^\beta)$ (and every qubit participates in $O(N^\beta)$ stabilizers) by taking a homological product of a larger number of codes, then for some sufficiently small $\beta$ this would imply code with record breaking properties of distance and rate.
A fourth application is to implementing logical Clifford operations on existing high distance codes.  While, for example, the toric code with logical qubits encoded using punctures has a well-defined methodology to implement logical operations by measuring products of logical qubits by operators which encircle multiple punctures, we do not have a general procedure to do logical Clifford operations on arbitrary codes.  We will see that weight reduction allows this.

From one point of view, it is perhaps not surprising that some kind of weight reducing technique should exist.  There is a general procedure to turn quantum codes into manifolds, first hinted at in  \cite{bh} and then later fully developed in \cite{fh}.  This procedure uses the stabilizers of the quantum code to define a chain complex, and then reverse engineers some scheme for attaching handles using that chain complex.  This gives a cell decomposition of some manifold, and the manifold can be given a metric.  If the quantum code is LDPC, then the resulting cell decomposition has a bounded geometry: every cell attaches to $O(1)$ other cells.  If the quantum code is not LDPC, then every code attaches to many cells; however, one might expect that one could refine the cellulation, decomposing each cell into many smaller cells, so that every cell only attaches to $O(1)$ other cells, and then define an LDPC quantum code from the resulting refinement of the cellulation.  One might hope that if the initial code is close to LDPC (so that, for example, the power $\beta$ in the above paragraph is close to $0$), then the qubit overhead entailed by the refinement would not be too large.

However, to put this abstract idea into practice encounters some obstacles.  For one thing, the procedure to turn a code into a manifold requires a ``lifting"\cite{fh} of the code to a chain complex over $\mZ$, and we do not know how such a lifting might change the appropriate norm of the rows and columns of the boundary operator: a code that is ``almost LDPC" might lift to some very non-sparse chain complex.  For another thing, this idea is fairly abstract, and for purposes of weight reducing a code, we would like a concrete procedure: the refinement needed would depend greatly on the geometry used to attach the various cells in the procedure of \cite{fh}.  Finally, we would like to do the weight reduction in as optimal a way as possible.

At the same time, there is one key factor working in our favor, namely, while we can draw inspiration from geometric ideas like ``refining a cellulation", we have no obligation to do the weight reduction in a way that makes sense topologically.  Roughly speaking, all we care about is that our various manipulations do the right thing homologically on certain chain complexes, without any need that those chain complexes come from cellulations of a manifold.

A previous paper that I wrote on weight reduction\cite{owr} claimed to give a procedure for weight reduction, but Z\"{e}mor has pointed out an error in one of the lemmas.  This paper corrects the error in that lemma; other lemmas in that paper are correct, including distance balancing and the operation described here as thickening.  With the error corrected, the techniques of that paper do not suffice to do weight reduction, but we introduce a new technique that makes weight reduction possible.  Using this technique and a more careful analysis, we further find that we can apply weight reduction to a very simple code, a code whose $X$-stabilizers are low weight, being chosen as the $X$-stabilizers of a classical (logarithmic) density parity check code, and whose $Z$-stabilizers are high weight, being chosen uniformly at random from operators that commute with the given $X$-stabilizers, to obtain an LDPC code with distance $\tilde \Omega(N^{2/3})$ and $\tilde \Omega(N^{2/3})$ logical qubits.  By tensoring our code with hypergraph product codes\cite{TZ14} (i.e., by distance ``unbalancing" and then balancing), we can attain (up to polylogs) the scaling of distance and rate given above for any $\alpha\in [2/3,1]$.

\subsection{Notation and Definitions}
Throughout when we say quantum code, we mean a CSS code.

Let us define some parameters of a quantum code.
We consider a quantum code with $N$ qubits, and $N_X,N_Z$ $X$- and $Z$-type stabilizers; when we refer to ``stabilizers" throughout, we mean generators of the stabilizer group.  Let $K$ be the number of logical qubits.

Assume each qubit participates in at most $q_X$ $X$-stabilizers and at most $q_Z$ $Z$-stabilizers.
Let $w_X$ denote the maximum weight of an $X$-stabilizer, i.e., the maximum number of Pauli operators in such a stabilizer.
Let $w_Z$ denote the maximum weight of a $Z$-stabilizer.  Let $d_X,d_Z$ denote the $X$ and $Z$ distances of the code, i.e., the minimal weight of a nontrivial $X$-type and $Z$-type logical operator.  

We will assume that all parameters $w_X,q_X,w_Z,q_Z$ are $\geq 1$, as if any of these parameters is $0$ then we have a classical code.

In the language of chain complexes, we have a  $2$-complex with three vector spaces $\cA_2,\cA_1,\cA_0$ over $\mathbb{F}_2$, with preferred bases, with basis elements (called cells) corresponding to $Z$-stabilizers, qubits, and $X$-stabilizers, and with boundary operators $\partial_1,\partial_2$ where $\partial_j$ is a linear map from $\cA_j$ to $\cA_{j-1}$.
All chain complexes in this paper are over $F_2$; we use the term $2$-complex to indicate that it has $0$-, $1$-, $2$-cells, but
this is a purely algebraic construction, rather than topological; no assumption is made, for example, that a $1$-cell has two $0$-cells in it boundary.

We sometimes say that a $1$-chain (we do not distinguish between chains and cochains) corresponds to a certain $Z$-type or $X$-type operator.  This is done in the obvious way: it is the product of Pauli $Z$ or $X$ on qubits which whose coefficient in that chain is equal to $1$.  We will freely switch between the terminology of chains and operators, depending on what is most natural.

The column weight of $\partial_1$ is bounded by $q_X$ and the row weight is bounded by $w_X$.  Similarly,
the row weight of $\partial_2$ is bounded by $q_Z$ and the column weight is bounded by $w_Z$.
Every nontrivial representative of $H_1$ has weight at least $d_Z$ and every nontrivial representative of $H^1$ has weight at least $d_X$.

The construction is simplest in the particular case of a code that has the property of being {\it reasonable}.  
We say that a code is reasonable if there is no $Z$-logical operator whose support is contained in the support of some $Z$-stabilizer.
So, if the $Z$ distance is greater than or equal to the largest stabilizer weight, then the code is reasonable.

Let $\supp(\cdot)$ denote the support of an operator.

If the code has the property that for every $Z$-stabilizer $S_Z$ and every $X$-stabilizer $S_X$,
the cardinality of $\supp(S_X) \cap \supp(S_Z)$ is equal to $0$ or $2$, then
the property of being reaosnable can be related to a certain graphical property (note, the property that 
$|\supp(S_X) \cap \supp(S_Z)|$ equals $0$ or $2$ holds automatically if every $X$-stabilizer has weight at most $3$ as the cardinality of the intersection must be even). 
To define this graphical property, for every $Z$-stabilizer $S$ with support $\supp(S)=Q$, define
a graph $G_Q$ with vertices corresponding to qubits in $Q$ and 
with an edge between vertices if they are both in the some $X$-stabilizer.
Then, a code is {\it reasonable} if, for every $Z$-type stabilizer $S$, the graph $G_{\supp(S)}$ has the property that, for every connected component of the graph, the product of Pauli $Z$-operators over qubits corresponding to vertices of the component is in the stabilizer group.

A code has the stronger property of being {\it connected} if, for every
$Z$-type stabilizer $S$, $G_{\supp(S)}$ is connected.

Given a reasonable code, we can choose different stabilizers (i.e., different generators of the same stabilizer group) to make the code connected: replace each stabilizer $S$ with one stabilizer for each connected component of $G_{\supp(S)}$.  Then, remove redundant stabilizers $S$ for which $G_{\supp(S)}$ is not connected.  Indeed, this can only help reduce $w_Z$; this cannot increase $w_Z$ or $q_Z$.
For example, if a code on two qubits has stabilizers $Z_1$ and $Z_1 Z_2$, with no $X$-type stabilizers, it is reasonable but not connected; however, we can replace the stabilizers with $Z_1$ and $Z_2$ to get a reasonable code with the same stabilizer group.

There is one interesting possible application of ``unreasonable" codes.  Suppose one has some code encoding several logical qubits.  Then suppose one wishes to impose some additional constraint on those logical qubits, say $\tilde Z_1 \tilde Z_2$, without separately imposing $\tilde Z_1$ or $\tilde Z_2$, and one wishes to do it by adding low weight stabilizers rather than directly adding some high weight logical operator.  This might be useful for preparing some interesting resource state.  We discuss this further later.

\subsection{Outline}
We are going to give a ``toolkit" for weight reduction, comprised of several different ways of transforming a code.
The first two we call ``copying" and ``gauging".  
Despite our use of the term ``gauging", this has nothing to do with gauge qubits in subsystem codes; it does involve introducing extra physical qubits to split stabilizers, but all codes are stabilizer, not subsystem.
We handle these two together, as implementing them in the same step can reduce overhead.
The others are called ``thickening" and ``coning".

Copying and gauging are, at their heart, operations on a classical code.  We illustrate them by their action on the $X$-stabilizers, but of course they can be done dually on the $Z$-stabilizers.  They can reduce $q_X,w_X$ to $O(1)$, but they unfortunately increase $w_Z$, and may impact the distance.
The gauging operation was in lemma 1 of \cite{owr}; this was the lemma with an error in it, corrected here.

Thickening was already introduced in \cite{owr}, under the name $Z$-type qubit splitting.  This is an operation that reduces $q_Z$ (or, dually, $q_X$).
This has the advantage that it reduces $q_Z$ without impacting the other parameters $w_Z,w_X,q_X$(to be precise, a large $q_X$ on the code input to thickening can lead to a large $w_Z$ on the output code, but otherwise the impact on other parameters is minimal).
Indeed, thickening {\it increases} $d_X$, though it does this at the cost of increasing $N$.

Coning is a new operation.  It can reduce $w_Z$, again with minimal impact on other parameters.
This coning operation is an essential part of weight reduction.  It is the main conceptual contribution of the
weight reduction of this paper.  This general construction
uses
two new key ideas, first the mapping cone from topology as a way to effectively induce stabilizers, and second, a way of adding additional relations to kill unwanted homology using recent results on cycle bases in graphs.  
We explain an example using the toric code where this coning process reduces to a familiar geometric construction of ``filling in holes".  
The mapping cone in a sense gives us a way to ``fill in holes" even when there is no obvious geometry for general codes.

Coning is a general procedure to weight reduce $Z$-stabilizers, but this can also be thought of as adding additional ``nonlocal" operators to a code.  That is, suppose one has some code that is already LDPC, and one wishes to add a stabilizer that is one of the logical operators of the code (thus, forcing a logical qubit to take a certain value); coning gives one a procedure to do this by adding some extra qubits and low weight stabilizers.  Thus, this may have some application in measuring logical operators in arbitrary LDPC quantum codes; we leave this for future work.

In \cref{together}, we will show that given these operations, any code can be weight reduced to an LDPC, with an increase in $N$ and decrease in distance by a factor polynomial in $w_X,w_Z,q_X,q_Z$.
However, we give all the tools here, as it seems difficult to say in advance what may be the most effective way to weight reduce a given code.  One might use the various tools in different sequences.

In \cref{applic}, we apply this weight reduction procedure to construct an LDPC code with distance $\tilde \Omega(N^{2/3})$.

In \cref{impsound}, we show how to improve the ``coning" construction, so that one of the important parameters of that construction (called ``soundness") can be improved, which in turns can improve the distance of the resulting code.  This is not needed for \cref{applic}, but is given if it may be useful later.  Further, as we remark in \cref{impsound}, the improving soundness procedure allows a partial derandomization of the construction of \cref{applic}.

\section{Copying and Gauging}
Copying is a simple technique to reduce $q_X$ (or, dually, $q_Z$).
We sketch this briefly, since we will give a detailed explanation later when combining it with gauging.
Simply concatenate a code with a repetition code on $q_X$ qubits in the $X$-basis, where the repetition code has stabilizers $X_1 X_2, X_2 X_3, \ldots, X_{q_X-1} X_{q_X}$.
This increases the $Z$-distance of the concatenated code, while leaving the $X$-distance unchanged, and increases the number of qubits.  Unfortunately, it increases the $Z$-stabilizer weight, because $Z$-operators of the original code get replaced by logical $Z$-operators of the repetition code, i.e., products of $Z$ over $q_X$ qubits.   However, it can reduce $q_X$: for each qubit of the original code, for each $X$-stabilizer acting on that qubit, we can choose a distinct representative of the logical $X$-operator.  In this case, a representative of logical $X$ in the repetition code is simply $X$ on a single qubit.
Thus, this reduces $q_X$ to at most $3$: one stabilizer of the original code, plus at most two stabilizers of the repetition code.

By combining copying and ``gauging" we can get a new code with $w_X,q_X \leq 3$.

We first explain the effect of copying and gauging on the $X$-stabilizers in \cref{nXs}, and then explain how to change the $Z$-stabilizers in \cref{nZs}.
The operation on the $X$-stabilizers is essentially the same as weight reducing a classical code\cite{hho}.
The operation on the $X$-stabilizers will increase the number of qubits and increase the weight of the $Z$-stabilizers.

Throughout this section, the ``original code" is the code before weight reducing, and the ``$X$-reduced code" is the code constructed in this section.  We will then in subsequent sections apply additional operations to the $X$-reduced code to get a final weight reduced code.

\subsection{New $X$-Stabilizers}
\label{nXs}
The procedure to reduce $q_X$ can be understood intuitively as making several copies of each qubit in the $X$-basis, adding $X$-stabilizers to enforce that they are copies,
and having different copies participate in different $X$-stabilizers,, while the procedure to reduce $w_X$ can be understood as the dual.

For each qubit $q$ of the original code, the $X$-reduced code will have $q_X$ qubits, labeled by a pair
$(q,j)$ where $j\in \{1,\ldots,q_X\}$.
Similarly, for each $X$-stabilizer $s$ of the original code, if the stabilizer acts on $d_s\leq w_X$ bits, the $X$-reduced code will have $d_s$ $X$-stabilizers, labelled by a pair
$(s,k)$ where $k\in \{1,\ldots,d_s\}$.
We add also
 $q_X-1$ additional $X$-stabilizers for each qubit $q$; label these additional stabilizers by a pair $[q,j]$ for $j\in \{1,\ldots,q_X-1\}$.
Further, add $d_s-1$ additional qubits for each $X$-stabilizer $s$; label these additional qubits by a pair $[s,k]$.

We will call qubits and $X$-stabilizers labelled by pairs $[s,k]$ and $[q,j]$ the ``new" qubits and stabilizers.  The other qubits and $X$-stabilizers are labelled by pairs $(q,j)$ and $(s,k)$, and are called the ``copied" qubits and stabilizers.
 We say that $(q,j)$ and $[q,j]$ are ``associated" with $q$ and $(s,k)$ and $[s,k]$ are ``associated" with $s$.

Thus, if the original code has $N$ qubits and $N_X$ $X$-stabilizers, the $X$-reduced code has
$O(N+N q_X)=O(N q_X)$ qubits and $O(N+N q_X)=O(N q_X)$ $X$-stabilizers. 
 
 We now specify the $X$-stabilizers.
 Each stabilizer $[q,j]$ is a product $$X_{(q,j)} X_{(q,j+1)}.$$
Each stabilizer $(s,k)$ is as follows.  Suppose stabilizer $s$ was a product $X_{q_1} X_{q_2} \ldots X_{q_{d_s}}$.
Then, we have
\begin{eqnarray}
&(s,1)=&X_{[s,1]} X_{(q_1,j_{s,1})} \\ \nonumber
1<k<d_s \quad \rightarrow \quad &(s,k)=&X_{[s,k-1]} X_{[s,k]} X_{(q_k,j_{s,k})} \\ \nonumber
&(s,d_s)=& X_{[s,d_s-1]} X_{(q_{d_s},j_{s,d_s})}.
\end{eqnarray}
 We choose the sequences $j_{s,k}$ so that each qubit is in at most $3$ stabilizers: at most $2$ new stabilizers, plus one other.
 
 We can explain this in the language of boundaries also.
 The stabilizer $[q,j]$ has qubits $(q,j)$ and $(q,j+1)$ in its coboundary.
Similarly, the qubit $[s,k]$ has boundary $(s,k)+(s,k+1)$ for each $k$.
Finally, for each qubit $q$ and each stabilizer $s$, such that $s$ was in the boundary of $q$, we pick some $j\in \{1,\ldots,q_X\}$ and some $k \in \{1,\ldots,d_s\}$ and then $(s,k)$ will be in the boundary of $(q,j)$, choosing $j,k$ so that each qubit has at most $3$ stabilizers in its boundary and each stabilizer has at most $3$ qubits in its coboundary.

\subsection{New $Z$-Stabilizers}
\label{nZs}
We now explain how to choose the $Z$-stabilizers of the $X$-reduced code.
For each $Z$-stabilizer $S$ of the original code, there will be some $Z$-stabilizer $\tilde S$ of the $X$-reduced code, and
these will be the only $Z$-stabilizers of the $X$-reduced code.

We now specify $\tilde S$.
Let $S$ be some given $Z$-stabilizer of the original code, with $S=Z_{q_1} Z_{q_2} \ldots Z_{q_w}$ for some $w$.
For every qubit $q$ of the original code, define
\be
\tilde Z_q=\prod_{j=1}^{q_X} Z_{(q,j)}.
\ee
To motivate this, recall that we have, in the previous subsection, ``copied" each qubit in the $X$-basis, and this $\tilde Z$ is a logical operator of a repetition code.

Let
\be
\tilde S=\tilde Z_{q_1} \tilde Z_{q_2} \ldots \tilde Z_{q_w} G,
\ee
where we next define the operator $G$ so that $\tilde S$ commutes with all stabilizers.

We now define $G$.
Each qubit $Q_a$ that $S$ acts on may in turn also be in some $X$-stabilizers.  Let $X(S)$ denote the set of $X$ stabilizers which are in the boundary of some qubit which is in the boundary of $S$.
Let
$$G=\prod_{s\in X(S)} G_s,$$
where $G_s$ is as follows.
Since $s$ commutes with $S$ by assumption, there are an even number of qubits which are in the support of both $S$ and $s$.
Denote these qubits by $q_{s,1},q_{s,2},\ldots$
Each of these qubits is in the coboundary of some stabilizer of the $X$-reduced code which is associated with $s$.
Let $q_{s,a}$ be in the coboundary of stabilizer $(s,j_a)$.
Order these qubits so that $j_1<j_2<\ldots$.
For each pair of successive qubits, $q_{s,2n+1},q_{s,2n+1}$ for $n=0,1,\ldots$,
define a ``string" which is a product of operators $Z_{[s,j_{2n}]} Z_{[s,j_{2n}+1]} \ldots Z_{[s,j_{2n+1}-1]}$.
Let $G_s$ be the product of these strings.

Then, indeed, the $Z$-stabilizers of the $X$-reduced code commute with {\it all} the $X$-stabilizers.

\subsection{$X$-Reduced Code Properties}
Let parameters with a tilde (such as $\tilde w_X,\tilde d_X,\ldots$) denote the corresponding parameters of the $X$-reduced code.

The following lemma corrects lemma 1 of \cite{owr}.  In particular, item 5 corrects an error in that one.

Also, lemma 1 of \cite{owr} only did the ``gauging" part of this construction, not the copying.
We have combined them here: while copying does increase $w_Z$ (due to the need to replace $Z$ operators of the original code by higher weight logical $Z$ operators of the concatenated code), this weight increase is negligible compared to weight increases
that are caused by the gauging operations.  So, there is no reason not to do both at once.

\begin{lemma}
The $X$-reduced code has the following parameters:
\begin{itemize}
\item[1.] $\tilde N=O(N q_X)$.

\item[2.] $\tilde K=K$.

\item[3.] $\tilde w_X,\tilde q_X=O(1)$.

\item[4.] $\tilde q_Z\leq {\rm max}(q_Z,w_X q_Z)$.

\item[5.] $\tilde w_Z\leq q_Z q_X (1+w_X).$

\item[6.] $\tilde d_Z\geq d_Z q_X$.

\item[7.] $\tilde d_X \geq d_X \Omega(1/w_X)$.
\end{itemize}
\begin{proof}
\begin{itemize}
\item[1.] Shown previously.

\item[2.] We have increased the number of qubits but also increased the rank of the $X$-stabilizer group by the same amount.

\item[3.] Immediate from construction of the code.

\item[4.] For any copied qubit of the $X$-reduced code, a $Z$-stabilizer acts on that qubit iff the corresponding $Z$-stabilizer of the original code acts on the qubit of the original code associated to that copied qubit.  Hence, the maximum number of $Z$-stabilizers acting on those qubits is bounded by $q_Z$.
For any new qubit $[s,k]$ of the $X$-reduced code, a $Z$-stabilizer $\tilde S$ of the $X$-reduced code can only act on that qubit if
$s$ is in $X(S)$.  Hence, the maximum number of $Z$-stabilizers acting on those qubits is bounded by $w_X q_Z$.
Hence, $\tilde q_Z \leq {\rm max}(q_Z,w_X q_Z)$.

\item[5.] The weight of each $\tilde Z$ equals $q_X$, so the weight of
$\tilde Z_{q_1} \tilde Z_{q_2} \ldots \tilde Z_{q_w}$ is bounded by $w_Z q_X$.
The weight of $G$ is equal to the sum over $s\in X(S)$ of the weight of $G_s$, which in turn is bounded by $w_X$.
The cardinality of $X(S)$ is bounded by $w_Z q_X$.
So, the weight of $G$ is bounded by $w_Z q_X w_X$.
So, $\tilde w_Z\leq w_Z q_X (1+w_X).$

\item[6.] 
It is useful to define two additional stabilizer groups.  First, consider the subgroup of the $X$-stabilizer group (of the $X$-reduced code) consisting of all stabilizers supported on the copied qubits.  Call this group $G_S^X$, the subscript standing for ``shortened", in analogy to the idea of a shortened code in coding theory.  Second, consider the group $G_P^Z$ (the subscript stands for ``punctured") of products of $Z$-operators acting on the copied qubits obtained by ``forgetting" the action of $Z$-stabilizers on the new qubits, i.e., an operator is in $G_P^Z$ iff it can be multiplied some product of $Z$-operators on the new qubits such that the product is in the $Z$-stabilizer group.
Note that $G_P^Z$ and $G_S^X$ commute with each other; further, they are the stabilizer group of the original
code concatenated with a repetition code in the $X$-basis.
Consider any $Z$-logical operator $L$.  This operator is a product of two operators $L_C L_N$ where
$L_C$ acts only on the copied qubits and $L_N$
acts only on the new qubits.
If $L_C$ is the identity, then $L_N$ must be also.
Note that $L_C$ must commute with $G_S^X$.  Further, we can multiply $L_C$ by any element of $G_P^Z$, changing $L_N$ in some way,
and the result is a multiplication of $L$ by an element of the stabilizer group.
Hence, if $L$ is nontrivial, then $L_C$ must be a nontrivial logical operator of the concatenated code.
Hence, the weight of $\tilde L$ must be at least $d_Z q_X$.

\item[7.] Consider any nontrivial $Z$-logical operator.  This operator is a product of two operators $L_C L_N$ where
$L_C$ acts only on the copied qubits and $L_N$
acts only on the new qubits.

If $L_N$ is the identity, it is easy to see that the weight of $L_C$ must be at least $d_X$.  We can multiply $L$ by an element of $G_S^X$, which leaves $L_N$ equal to the identity but changes $L_C$.  The operator $L_C$ must commute with $G_P^Z$.  Hence, $L_C$ must be a nontrivial logical operator of the concatenated code and so must have weight at least $d_X$.

Suppose instead $L_N$ is not equal to the identity.  However, by multiplying $L$ by a product of copied stabilizers, we can turn $L_N$ into the identity, increasing the weight of $L$ by a factor of at most $w_X$.
\end{itemize}
\end{proof}
\end{lemma}

\section{Thickening}
\label{thicken}
We now explain how to reduce $q_Z$.  This is the same as in \cite{owr}, and we mostly reproduce the results there, with some
added discussion.

\subsection{Thickening Construction}
This step is done in two substeps.  
In the first step, we take some code $C$ (in this case, $C$ is the code obtained from the previous section after reducing the $X$-stabilizers).  We write that code as a chain complex $\cC$, associating $X$-stabilizers with $0$-cells, qubits with $1$-cells, and $Z$-stabilizers with $2$-cells (this shifts the dimension by $1$ from the convention used in \cite{owr}, where qubits were associated with $2$-cells).  We take the homological product of this complex $\cC$ with a chain complex $\cE$
corresponding to a cellulation of an interval; $\cE$ has only $0$-cells and $1$-cells in its chain complex.
This step may be geometrically interpreted as ``thickening" the complex, by taking its product with an interval.  The resulting product complex is called $\cD$ and the product code is called $D$.

In the second step, we observe that there is a large redundancy among the $Z$-type stabilizers of $D$: $Z$-type stabilizers at different values of the coordinate in the direction of the interval differ only by a product of other stabilizers.  
Let us call this coordinate in the interval direction the ``height".
This can be understood as a result of the homological product: the chain complex $\cE$ has $1$-cells, and as a result the product $\cC \otimes \cE$ has $3$-cells which encode redundancies among the $Z$-stabilizers of $D$.
This allows us to remove many of the stabilizers in $D$ and obtain a code $\tilde C$ with smaller $q_Z$.  Geometrically, one may view this as follows: since the complex has been thickened, we can attach the cells corresponding to stabilizers at different heights, to avoid attaching too many to any given cell.

To define $\cE$, fix an integer integer, $\ell>1$.
Define chain complex $\cE_{1} \stackrel{\partial^\cE_1}{\rightarrow} \cE_0$,
where ${\rm dim}(\cE_0)=\ell$ and ${\rm dim}(\cE_1)=\ell-1$, with
\be
\partial^\cE_1=\begin{pmatrix}
1 & \\
1 & 1  \\
  & 1 & 1 \\
 & & & . & . \\
& & & & . & . \\
& & & & & . & . \\
& &  & & & & 1 & 1\\
& &  & & & &  & 1
\end{pmatrix}.
\ee
This chain complex can be interpreted geometrically as a cellulation of an interval, with $\ell$ $0$-cells and $(\ell-1)$ $1$-cells.
We have $b_1(E)=0,b_0(E)=1$.

By the  K\"{u}nneth formula, this code $D$ has the same number of logical qubits as $C$ does.
This code has $Nl+n_X(\ell-1)$ qubits and $\ell n_X$ $X$-type stabilizers.

We now define the code $\tilde C$.  
This code $\tilde C$ will have the same number of qubits and $X$-type stabilizers as $D$ does.
This code will be obtained by taking a subset of the $Z$-type stabilizers of the stabilizers of $D$, while taking all of the $X$-type stabilizers.  What we will show is that although we take only a subset of the $Z$-type stabilizers,
the code $\tilde C$ will have the same stabilizer group as $D$.

Recall that the $Z$-type stablizers are in one-to-one correspondence with basis elements of $\cD_2=\cC_2 \otimes \cE_0 \oplus \cC_1 \otimes \cE_1$.  We keep all $Z$-type stabilizers corresponding to basis elements of $\cC_1 \otimes \cE_1$.  However, for each,  basis elements of $\cC_2$,
we keep only one $Z$-type stabilizer in $\cC_2 \otimes \cE_0$.  Let $w_1,\ldots,w_\ell$ be basis vectors for $\cE_0$ in the standard basis.
For each basis element $v\in \cC_2$ in the standard basis, we pick one integer $k$, with $1 \leq k \leq \ell$, and we keep the stabilizer corresponding to $v \otimes w_k$.

In the following lemma, $\tilde K$ refers to the number of logical qubits of $\tilde C$, while various $\tilde w_X,\tilde w_Z,\tilde q_X,\tilde q_Z$ refer to the appropriate parameters of $\tilde C$.

Note item 7, showing that the distance increases by a factor of $\ell$.  This increase in distance is the same as that from distance balancing in \cite{owr}. 

Remark: this thickening technique is essentially the same as the distance balancing in \cite{owr}, using an extra trick of ``choosing heights" to reduce $q_Z$.  It improves $X$-distance while increasing number of physical qubits and worsening rate since the number of logical qubits in unchanged.  So, the reader might wonder whether we can do thickening in a way more similar to \cite{ekz}, that would
reduce $q_Z$ while leaving the rate unchanged.
We do not see any way to do this, however: the reduction in $q_Z$ depends on a particular form of redundancy among the $Z$-stabilizers which would not be present in the distance balancing of \cite{ekz}.  However, we leave this as an open problem.

\begin{lemma}
\label{prodC}
The code $\tilde C$ has the following properties:
\begin{itemize}
\item[1.] All stabilizers commute with each other.  $\tilde N= Nl+n_X(\ell-1)$ and  $\tilde n_Z=n_Z+(\ell-1)N$ and $\tilde n_X=\ell n_X$.  

\item[2.] Codes $\tilde C$ and $D$ have the same stabilizer group.
Explicitly, for each $Z$-type stabilizer of the form $Z_{q_1} Z_{q_2} \ldots Z_{q_w}$ in $C$,
the stabilizer group  of $\tilde C$ contains
$Z_{(q_1,m)} Z_{(q_2,m)} \ldots Z_{(q_w,m)}$ for all $1 \leq m \leq \ell$.

\item[3.] $\tilde  K =K$.

\item[4.] $\tilde w_X=w_X+2$ if $\ell\geq 3$ and $\tilde w_X=w_X+1$ if $\ell=2$.

\item[5.] $\tilde w_Z={\rm max}(w_Z,2+q_X)$.

\item[6.] $\tilde q_X={\rm max}(q_X,2)$.

\item[7.] $\tilde d_X=\ell d_X$.

\item[8.] $\tilde d_Z=d_Z$.
\end{itemize}
\begin{proof}
This is lemma 2 in \cite{owr}.
\end{proof}
\end{lemma}

\subsection{Choosing Heights}
The value of $\tilde q_Z$ will depend on which stabilizers $Z_{(q_1,k)} Z_{(q_2,k)} \ldots Z_{(q_w,k)}$  we choose to include.  If we keep the same $k$ for all $Z$-type stabilizers of $C$, then we have no improvement in $\tilde q_Z$.  However, if we choose different $k$ for different stabilizers then we can reduce $\tilde q_Z$ by making different stabilizers act on different qubits $(q,k)$.  As shown in \cite{owr}, for any desired $\tilde q_Z$, we can attain a code with that $\tilde q_Z$ by choosing large enough $\ell$.
The following lemma gives upper bounds on how big an $\ell$ is needed.

\begin{lemma}
\label{qsplitlemma2}
Let $w$ be any positive integer.
For $\ell$ sufficiently large that
\be
2e {q_Z \choose w+1} (\frac{1}{\ell})^{w+1} {\rm min}(q_Z w_Z,N) \ell \leq 1,
\ee
there is a choice of $k$ for each $Z$-type stabilizer of $C$ such that
\be
\tilde q_Z \leq {\rm max}(w+2,w_X).
\ee
\begin{proof}
This is lemma 3 in \cite{owr}.
\end{proof}
\end{lemma}

The relation between $\tilde q_Z$ and $w$ is as follows: the lemma shows that there is a choice of $k$ for each 
$Z$-type stabilizer of $C$
such that each qubit corresponding to basis element of
$\cC_1 \otimes \cE_0$ is in at most $w$ different stabilizers corresponding to basis elements of
$\cC_2 \otimes \cE_0$.
Additionally, each qubit a corresponding to basis element of
$\cC_1 \otimes \cE_0$ is in at most $2$ stabilizers corresponding to basis elements of
$\cC_1 \otimes cE_1$.
There are also qubits corresponding to basis elements of $\cC_0 \otimes \cE_1$.
These are in at most $w_X$ stabilizers corresponding to basis elements of $\cC_1 \otimes \cE_1$.

Let us make some comments on this lemma.
First, note that one might have guessed that it would suffices to take $\ell$ proportional to $q_Z$ in order to attain $\tilde q_Z=O(1)$.  The reasoning behind this would be, roughly: ``if each qubits participates in at most $q_Z$ different stabilizers, then if I thicken by an amount $\ell \sim q_Z$, I can put each stabilizer at a different position."
Unfortunately, the lemma is slightly weaker than this, and this rough reasoning is not correct.

However, the lemma is not much weaker.  Fix any $\tilde q_Z=O(1)$.  Assume that the code $C$ that is input to this procedure has already been copied and gauged so that $w_X=3$.  So, one may take $$\ell=O(q_Z^{\frac{w+1}{w}} {\rm min}(q_Z w_Z,N)^{1/w}),$$
with $w=\tilde q_Z-2$.
Hence, we have
\begin{lemma}
\label{Oeps}
For any $\epsilon>0$, we can pick $\tilde q_Z=O(1)$ with
$\ell=O(q_Z^{1+\epsilon}  {\rm min}(q_Z w_Z,N)^{O(\epsilon)}).$
\end{lemma}

Also we have
\begin{lemma}
\label{hgc}
We can pick $\tilde q_Z=O(1)$ with $\ell=q_Z w_Z+1$.
\begin{proof}
This follows from a relationship with graph coloring.  Considering a graph with vertices corresponding to stabilizers, with two vertices neighboring if share a qubit in their support.  This graph has degree $\leq q_Z w_Z$ and so can be colored with at most $q_Z w_Z+1$ colors. 
\end{proof}
\end{lemma}

The proof of lemma \ref{qsplitlemma2} is probabilistic.  It relies on the Lovasz local lemma.  If one is willing to accept a constant factor increase in $\ell$, then there are efficient algorithms to find the needed choice\cite{moser2010constructive}.

Also, if one is will to work with the slightly weaker result that such a choice of $k$ for each $Z$-type stabilizer
exists whenever
${q_Z \choose w+1} (\frac{1}{\ell})^{w+1} N \ell<O(1),$ then the proof of the lemma shows that choosing the $k$ independently and uniformly at random will work with probability $\Omega(1)$.

We also remark that lemma \ref{qsplitlemma2} can be understood in terms of a variant of a standard problem in graph theory, called
``strong hypergraph coloring"\cite{agn}.
In graph coloring, one has a graph, and one must color each vertex from some number of colors so that for every edge, both vertices attached to that edge have different colors.
In hypergraph coloring, a hyperedge may have more than two vertices, and so there are several possible coloring problems.
In strong hypergraph coloring, the requirement is that {\it every} vertex in a hyperedge must have different colors.

To relate these, consider a hypergraph, where hyperedges correspond to qubits of $C$ and vertices correspond to $Z$-stabilizers of $C$.
The vertices in each hyperedge are the set of $Z$-stabilizers acting on that qubit.  Then, a coloring with at most $\ell$ different colors corresponds to a choice of $k$ for each $Z$-stabilizer of $C$.  The case $w=1$ corresponds to a strong hypergraph coloring.
It is possible\cite{strongbound}
that such a strong hypergraph coloring can require a number of colors of order the vertex degree times the hyperedge degree, i.e., $\ell=\Theta(w_Z q_Z)$.  However, what we see is that from lemma \ref{Oeps} is that by considering colorings in which, for every hyperedge, no color in that hyperedge is repeated more than $w$ times, we can use a number of colors that is closer to $q_Z$.

\section{Coning}
We mostly consider the case of reasonable codes as that is the simplest and most important one.  In \cref{unreasonable}, we consider the case of ``unreasonable" codes.

In \cref{geometric}, we give an illustration of the procedure here in a simple case related to the toric code.
This may help clarify it.
We also give
some geometric interpretation.

At this point, we need some definitions from homological algebra.
Throughout this section, all chain complexes will be over $\mathbb{F}_2$.
First we need:
\begin{definition}
Given two chain complexes $\cA,\cB$, with corresponding boundary operators $\partial^\cA,\partial^\cB$, a linear map $f:\cB \rightarrow \cA$ is called a chain map if it commutes with the boundary:
$\partial^\cA \circ f = f \circ \partial^\cB$.
\end{definition}

We recall the definition of the mapping cone of a chain map.  Here the chain map is assumed to preserve degrees, so that it
maps $\cB_j$ to $\cA_j$.
\begin{definition}
Given a chain map $f:\cB\rightarrow \cA$, the {\it cone} of $f$, written $\Cn(f)$, is the following chain complex.
The vector space $\Cn(f)_j$ is defined to be $\cA_{j} \oplus \cB_{j-1}$, and the boundary operator is defined by
\be
\label{conedef}
\partial\equiv \begin{pmatrix} \partial^\cA & f \\ 0 & \partial_B \end{pmatrix}.
\ee
\end{definition}

One useful example of a cone is when $\cA=\cB$ and the map $f$ is the identity map.  In this case, all homology groups of $\Cn(f)$ are trivial.  
To see this, consider some $j$-chain $u=x_j\oplus y_j$ where $x_j\in \cA_{j}$ and $y_j\in \cB_{j-1}$.
Then if $\partial u=0$, we have $\partial y_j=0$ and $\partial x_j=y_j$.  This implies that $u$ is a boundary: it is the boundary of the $(j+1)$-chain given by $0 \oplus x_j$.
This is a special case of a more general result: if the map $f$ induces an isomorphism on all homology groups, then
all homology groups of $\Cn(f)$ are trivial\cite{ncl}.

We now describe the construction to reduce $Z$-stabilizer weight using a cone.  
We first construct a code we call the ``cone code"; this code will have reduced $Z$-stabilizer weight but unfortunately doing this may create large $X$-stabilizer weight and $q_X$  While this may seem to merely delay the problem, the particular way in which the code code is constructed means that it will be easy to reduce the weight of those $X$-stabilizers by a process explained later.

Consider some quantum code $C$.  Let $C'$ be some code that contains the same qubits as $C$, and contains the same $X$-stabilizers
as $C$, but has only some subset of the $Z$-stabilizers.  These will, in most applications, be stabilizers of $C$ which are already low weight, and do not need to be weight reduced.  We will call these the ``direct stabilizers".
The coning construction will be used to ``induce" the remaining stabilizers.

Let $\cA$ be the chain complex corresponding to $C'$

We now construct a complex $\cB$.  The complex has $-1$-cells, $0$-cells, and $1$-cells.  We may think of $\cB$ as describing a classical code with the $0$-cells being checks, the $1$-cells being bits, and the $-1$-cells being ``redundancies" among the checks.
The complex $\cB$ is given by
$$\cB=\oplus \overline{\cB}_i.$$
Before defining the $\overline{\cB}_i$, we define some complexes $\cB_i$ from which the $\overline{\cB}_i$ are constructed.

The chain complex $\cB_i$ has $0$-cells and $1$-cells (no $-1$-cells).   The complex $\overline{\cB}_i$ defined later is given by adding additional $-1$-cells to $\cB$, while constructing some boundary operator $\partial_0$; this is done to kill
the zeroth 
homology of $\cB_i$.

{\bf Definition of $\cB_i$:}
Let $Q_1,Q_2,\ldots,Q_m$, for some integer $m$, be some sequence of sets of qubits of the given code.  
If we are using the code constructed after a previous ``thickening" step, we will often choose the $Q_i$ to be subsets of qubits corresponding to basis elements of $\cC_2 \otimes \cE_0$ rather than $\cC_1 \otimes \cE_1$, simply because there may be no need to weight reduce the other elements.

Choose these sets $Q_i$ such that, for every $i$, the product of Pauli $Z$ operators on all qubits in set $Q_i$ commutes with all $X$-stabilizers of the code.
The chain complex $\cB_i$ will have $1$-cells labeled by elements of $Q_i$.

For each set $Q_i$, define a set $S_i$ as follows.  Each $S_i$ will be the set of $X$-stabilizers of the quantum code which have support on at least one qubit in $Q_i$.
Each stabilizer in $S_i$ must in fact have support on an even number of qubits of $Q_i$ by assumption.  If we have weight reduced those stabilizers so that they have weight $3$, then indeed each stabilizer in $S_i$ must have support on two qubits of $Q_i$.
If we have not weight reduced in this way, then we need to make one more set of choice: for each stabilizer in $S_i$, ``pair off" the qubits in $Q_i$ on which that stabilizer has support in some way.  That is, if some stabilizer has support on $k$ qubits in $Q_i$, we find $k/2$ pairs of qubits, putting one qubit in each pair.  Then, $0$-cells of $\cB_i$ will be in one-to-one correspondence with a tuple consisting of, first, an element of $S_i$, and, second, one of these pairs; we write such a tuple $(S,j)$ where $S\in S_i$ and $j$ labels a pair.  Again, if stabilizers have weight $3$, the $0$-cells simply correspond to elements of $S_i$.
  
The boundary operator of the chain complex $\cB_i$ will be the obvious one: the boundary of a given $1$-cell is the sum of all tuples $(S,j)$ such that the corresponding qubit is in the given pair.
Again, if stabilizers have weight $3$, the boundary of a given $1$-cell is the sum of all $0$-cells
such that the corresponding qubit is in the corresponding stabilizer.

This chain complex then corresponds to a classical code, with bits corresponding to $1$-cells, and with checks corresponding to $0$-cells, such that every check acts on two bits.

If $\cB_i$ has trivial zeroth homology, then we take $\overline{\cB}_i=\cB_i$.
While the case that $\cB_i$ does not have trivial zeroth homology is the most interesting one, the reader may, on first reading, want
to skip the definition of $\overline{\cB}_i$ and simply assume that $\cB_i$ does have trivial zeroth homology, in order to understand in the simplest case how the coning construction gives the desired result.

Let us sketch what this ``desired result" is.
For each $i$, the complex $\cB_i$ has some first homology.  By assumption, the sum of all $1$-cells is an element of first homology, but there may be other elements.  Each such element of first homology corresponds to some subset of qubits of the code $C$.  The coning construction will ``effectively" induce, for each such subset, $Z$-type stabilizer given by the product of Pauli $Z$-operators on that subset.

Remark: assuming that the code is reasonable, we then choose, for each $Z$-stabilizer of the code, one set $Q_i$ equal to the support of that stabilizer.  If the code is not reasonable, then potentially this choice will not produce the original code but rather will increase the stabilizer group by adding some $Z$-logical operators to the stabilizer group.

{\bf Choice of Chain Map}
Define, for each $i$, a chain map $f_i:\cB_i\rightarrow \cA$ in the obvious way: it maps each $j$-cell in $\cB_i$ for $j\in \{0,1\}$ to the corresponding $j$-cell in $\cA_i$.
If we have, for some stabilizer in $S_i$, more than one pair, we map all tuples $(S,i)$ for $S\in S_i$, to the $0$-cell corresponding to $S$.

Define a chain map $\overline f_i:\overline{\cB}_i\rightarrow \cA=f_i \circ \Pi_i$, where $\Pi_i:\overline{\cB}_i \rightarrow \cB_i$ maps every $0$-cell or $1$-cell in $\overline \cB_i$ to the corresponding cell in $\cB_i$ and vanishes on $-1$-cells.

Finally, let $\cB=\oplus_i \overline{cB}_i$, and define a chain map $f:\cB \rightarrow \cA$ by
$$f=\begin{pmatrix} \overline f_i \\ \overline f_2 \\ \ldots \\ \overline f_m\end{pmatrix}.$$

\begin{definition}
The {\it cone code} for the given $Q_1,\ldots,Q_m$, and $S_1,\ldots,S_m$, $\overline{\cB}_i$, and $C$ is the code corresponding to $\Cn(f)$.
\end{definition}

We will compare this cone code to another code.  
\begin{definition}
The {\it induced code} for the given choices of sets $Q_i$, of pairings, and of $C$,
is a code with the same qubits and $X$-stabilizers as $C$, and with the set of $Z$-stabilizers given by the $Z$-stabilizers of $'$ as well as, for every element
of first homology of each $\cB_i$, the product of Pauli $Z$-operators over qubits corresponding to that element of first homology.
For each $i$, we say that the stabilizers {\it induced by $\cB_i$} are those given by elements of first homology of $\cB_i$.

We call the $Z$-stabilizers of the induced which correspond to $Z$-stabilizers of $C'$ the direct $Z$-stabilizers
and we call the other $Z$-stabilizers the induced $Z$-stabilizers.
\end{definition}

The cone code is not necessarily LDPC, but it may have some improved properties of $Z$-stabilizer weight, depending on $C$.  We later apply
a further transformation to the cone code, defining a ``reduced cone code", so that the result is LDPC assuming certain properties of $C$.

{\bf Construction of $\overline{\cB}_i$:}
We now construct the $\overline{\cB}_i$.  As mentioned, the reader may wish to skip this on first reading and proceed to lemma \ref{conecode}, taking every occurrence of $\overline{\cB}_i$ in that lemma to be $\cB_i$ and assuming that $\cB_i$ has trivial first homology.
For notational simplicity, fix some $i$, and let $\cF=\cB_i$.  We will construct some $\overline{\cF}$ and let $\overline{\cB}_i=\overline \cF$.  Write $S=S_i$ and $Q=Q_i$.

From $\cF$, construct a graph $G_i$ as follows.  Each edge corresponds to a $0$-cell in $\cB_i$ and
 each vertex corresponds to a $1$-cell in $\cB_i$.  An edge attaches two vertices when the corresponding $0$-cell is in the boundary of the two corresponding $1$-cells.

 The induced code can then be expressed very simply in terms of these graphs: for each connected component of the graph $G_i$, the product of Pauli $Z$ operators over all qubits in that component is a $Z$-stabilizer of the induced code, and those are the only stabilizers induced.  Further, for a reasonable code, these will all be elements of the stabilizer group, i.e., we do not induce any unwanted stabilizers.
  
 For notational simplicity, we will refer to $G_i$ simply as $G$ in this part of the construction, and refer to $Q_i$ as $Q$.

The following lemma guarantees the existence of pairings with certain properties.  We always make such a choice.
\begin{lemma}
\label{pairchoose}
For a reasonable code, for any pairing, the product of Pauli $Z$ operators in any connected component of $G$ is an element of the stabilizer group.
\begin{proof}
Suppose some connected component does not correspond to a $Z$ stabilizer.  Then, there must be some $X$ stabilizer which acts on an odd number of qubits in that component.  Then, for any pairing, there must be at least one edge leaving that component, so indeed it cannot be a connected component.
\end{proof}
\end{lemma}

Elements of zeroth cohomology of $\cF$ correspond to closed $1$-chains on this graph $G$.

By the decongestion lemma\cite{fh},
we can find a basis of simple cycles for the zeroth cohomology of $\cF$ which has weight $O(n \log(n))$, where $n=|Q_i|$, and where the weight of a cycle basis is the sum of the lengths of the cycles in that basis.
Further, each edge appears in the basis at most $O(\log(n)^2)$ times.

We construct $\overline\cB_i$ by adding a $-1$-cell to $\cB_i$ for each cycle in such a basis.  The coboundary of the given $-1$-cell is the sum of edges in the basis.

{\bf Geometric Interpretation of Added $-1$-cells:}
We can give a geometric interpretation of the added $-1$-cells.  Given graph $G_i$, we can extend that graph to a $2$-complex we call $\cS_i$ with trivial first homology by attaching a $2$-disc to each cycle in the basis.  Then, vertices, edges, and faces of that $2$-complex become $1$-, $0$-, and $-1$-cels of $\overline \cB_i$ respectively.  When constructing the cone code, these get shifted in dimension by $1$ to become $2$-, $1$, and $0$-cells.  Thus, in a sense the dimensions appear ``upside down": the $j$-cells of $\cS_i$ become $(2-j)$-cells in the cone code.
A crucial property for use later is that each $1$-cell in $\cS_i$ has exactly two $0$-cells in its boundary.

{\bf Reducing the Cone Code:}
There is no upper bound on the length of a cycle other than the trivial upper bound that it has length at most $|Q_i|$; indeed, if the graph $G$ is a cycle graph, then this is the length of a cycle on that graph.  This means that the coboundary of a $-1$-cell in $\overline \cB_i$ may have weight $|Q_i|$, and hence we may produce $X$-stabilizers of the cone code with weight up to $|w_Z|$ (of course, there may also already have been some high weight $X$-stabilizers in code the induced code).

Further, it is possible
that a given edge
in $G_i$ attaches to polylogarithmically many $2$-discs in $\cS_i$ if that edge appears in many cycles in the cycle basis.  This can lead to a $1$-cell in the cone code having polylogarithmically many $0$-cells in its boundary.

However, both these problems can be solved.  We describe the procedure to solve them.  The result we call the ``reduced cone code".

 To solve the second problem, let us say an ``added $j$-cell" of the cone code is a $j$-cell corresponding to $(j-1)$-cell in some $\overline \cB_i$.
Thicken the cone code (dually, i.e., interchanging $X$ and $Z$), and then choose heights for the added $0$-cells so that
at most one such added $0$-cell attaches to each $1$-cell.  Here we are using thickening only to reduce the number of added $0$-cells which attach to $1$-cells, and we are not concerned with the number of other $0$-cells attaching to a $1$-cell.
So, for any $\epsilon>0$, this can (by the same argument as \cref{Oeps}) be done by thickening by an
amount $\Theta(\log(|w_Z|)^{2+2\epsilon} |w_Z|^{\epsilon})$; that is, here $|w_Z|$ plays the role of $N$ in lemma \cref{Oeps}.

To solve the first problem, 
consider the subcomplex consisting just of a single added $0$-cell of the cone code, the added $1$-cells in its coboundary, and the added $2$-cells in their coboundary.  Call that subcomplex $\cD$.  Consider a single cycle in some graph $G_i$.  Consider the cycle graph containing just the edges in that cycle.  Regard that graph as a $1$-complex and attach a $2$-disc whose boundary is that cycle; call the result $\cE$.  Then, $\cD$ and $\cE$ are the same complex, up to a change in the dimension of cells, so that $j$-cells in $\cE$ corresponding to $(2-j)$-cells in $\cD$, and up to transposing the boundary operator.
However, it is easy to see geometrically how to make $\cE$ sparse, so that every cell in $\cE$ has only $O(1)$-cells in its boundary and coboundary: simply cellulate the added $2$-disc.  For example, labeling the vertices $0,1,2,\ldots,w-1$ in order, add an edge between vertices $j$ and $w-j$ for all $j$ with $0<j<w-j$.

\subsection{Comparison of Reduced Cone Code and Induced Code}
We now compare the reduced cone code and the induced code.

Finally, let us define a {\it soundness factor} $\lambda_i$.  Since each $\overline\cB_i$ has trivial zeroth homology, any closed $0$-chain $u\in \overline\cB_i$ equals $\partial v$ for some $1$-chain $v\in \overline\cB_i$.
Define $\lambda_i$ by
$$
\lambda_i={\rm min}_{u\in (\overline\cB_i)_0, \partial u=0, u\neq 0} \Bigl( {\rm max}_{v\in {\overline\cB_i}_1,u=\partial v} \frac{|u|}{|v'|}\Bigr),$$
where $|\cdot|$ denotes Hamming weight, and where $v'$ is the projection of $v$ into $\cB_i$.

The soundness factor $\lambda_i$ can be lower bounded in terms of the Cheeger constant of the graph $G_i$.
If a graph $G$ is connected,
let
$$
h(G)={\rm min}_{w,0<|w|\leq |V|/2} \frac{|\partial^T w|}{|w|},
$$
where the minimum is over all {\it vertex chains}\footnote{Here, by vertex chain, we mean a $0$-chain on the graph.  We use the term vertex chain because we want to refer to $0$-cells of the graph as vertices.} $w$   with Hamming weight $>0$ and at most $|V|/2$, where $V$ is the set of vertices of the graph $G$.  This is the usual definition of a Cheeger constant: the ratio is the familiar ratio of number of edges which are adjacent to one vertex in $w$ to the number of vertices in $w$, except we have used $\partial^T$ rather that $\partial$ (as is more common notation in graph theory) to be consistent with notation in terms of complexes.

If the graph $G$ is not connected, we define its Cheeger constant to be the minimum of the Cheeger constants of its connected components.

Then:
\begin{lemma}
\be
\lambda_i \geq h(G_i).
\ee
\begin{proof}
Let $v$ be some choice of $1$-chain in the definition of $\lambda_i$, with $v'$ the projection of $v$ on $\cB_i$.  
Without loss of generality, assuming $|v'|\leq |V_i|/2$, where $V_i$ is the vertex set of $G_i$; if $|v'|>|V_i|/2$, add the sum of all $1$-cells to $v$, which does not change $\partial v$.

Suppose for some edge of $G_i$, $v$ vanishes on one vertex attached to that edge, but is nonvanishing on the other vertex attached to that edge.  
Then
$\partial v$ equals $1$ on the $0$-cell corresponding to the given edge.  Let $w$ be the vertex chain on the graph corresponding to $v'$ so $|w|=|v'|$; then $|\partial^T w|$ is lower bounded by $h(G_i) |w|$ and $|u|\geq |\partial^T w|$.
\end{proof}
\end{lemma}

We have
\begin{lemma}
\label{conecode}
Let $\epsilon>0$.
Let $\ell=\Theta(\log(|w_Z|)^{2+2\epsilon} |w_Z|^{\epsilon})$.
Let parameters without a tilde denote parameters of the induced code.  Let parameters with a tilde denote parameters of the reduced cone code.
Let $w'_Z$ denote parameters of code $C'$.
Then,
\begin{itemize}
\item[1.] $\tilde N \leq O(N+\sum_i |Q_i| q_X) \ell$. 

\item[2.] $\tilde K=K$.

\item[3.] $\tilde w_Z\leq q_X+1+w'_Z+O(1)$.

\item[4.] $\tilde q_Z\leq q_Z+O(1)$.

\item[5.] $\tilde w_X={\rm max}(w_X,O(1))$.

\item[6.] $\tilde q_X={\rm max}(q_X,O(1))$.

\item[7.] $\tilde d_X\geq d_X$.

\item[8.] $\tilde d_Z\geq d_Z \lambda \ell$, where $\lambda={\rm min}(1,{\rm min}_i \lambda_i)\geq {\rm min}(1,{\rm min}_i h(G_i))$.
\end{itemize}
\begin{proof}
It is convenient to consider two more codes.  In each complex $\cS_i$, cellulate all the added $2$-discs at once, (using the same cellulation as done for each individual disc when defining the reduced cone code) to define some complex we call $\hat \cB_i$.
Then, define $\hat B=\oplus_i \hat \cB_i$ and define a code by coning with the obvious chain map from $\hat \cB$ to $\cA$.
Call the resulting code the celled code.  Thicken the celled code by the same amount as in defining the reduced cone code, and call the result the thick celled code.

Thus, one can think of the thick celled code as having, for every cellulated $2$-disc in the reduced cone code, many copies of that cellulated $2$-disc.
Let $d$ label various discs in the reduced cone code, and let $d_1,d_2,\ldots$ denote the corresponding copies of that disc in the thick celled code.  For any $1$-chain in the thick celled code, for any disc $d$, we can add boundaries or coboundaries to push the chain off all the cells in the interior of all but one copy $d_a$.  In this way, the chain can be turned into a chain that vanishes except on cells in the reduced cone code.
Hence the thick celled code has the same number of logical qubits as the reduced cone code and the distance of the reduced cone code is at least as large as the of the distance of the thick celled code, as every chain in the cone code defines in an obvious way a chain in the thick celled code.  In turn, the $X$-distance of the thick celled code is that same that of the celled code, while the $Z$-distance is increased proportional to the amount $\ell$ we thicken.

So, in all proofs of distance, we will prove distance of the celled code, and the result for the reduced cone code will follow.

For notational convenience, let $\cQ_i$ denote the space $(\cB_i)_1)$, let $\cX_i$ denote the space $(\cB_i)_0$.
Let $\hat{\cQ}_i$ denote the space $(\hat{\cB}_i)_1)$, let $\hat\cX_i$ denote the space $(\hat\cB_i)_0$,
and let $\hat{\cR}_i$ denote the space $(\hat\cB_i)_{-1}$, where $Q,X,R$ refer to ``qubit, $X$, relation".
Let $\cQ_i^\perp,\cX_i^\perp$ denote the subspaces of $\hat\cQ_i,\hat \cX_i$ which vanish on $\cQ_i,\cX_i$, so
$\hat\cQ_i=\cQ_i\oplus\cQ_i^\perp$ and similarly for $\hat\cX_i$.

Given a space such as $\cQ_i$, we write $\cQ[-1]$ to denote the space in $\Cn(f)$ corresponding to $\cQ$, and similarly for all other spaces defined in the paragraph above.  Here
the $-1$ in brackets denotes the shift in dimension:
for an arbitrary complex $\cD$, define the complex $\cD[-1]$ to have its $(j+1)$-cells in one-to-one correspondence with
$j$-cells of $\cD$, and with the boundary operator defined in the obvious way.

We have:
\begin{itemize}
\item[1.] 
There are $N$ $1$-cells in $\cA$.  Each graph $G_i$ has at most $|Q_i| |q_X|$ edges giving a total of 
$N+\sum_i |Q_i| q_x)$ $1$-cells, which gets multiplied by $\ell$ after thickening.  We must also account for cells in the interior of each attached disc.  This weight is at most $O(|Q_i| \log(|Q_i|))$ and does not get multiplied by $\ell$ because every such disc gets attached at only one height.

\item[2.] This follows from the proof of item 7 below.

\item[3.] Each $Z$-stabilizer in $\cB_i$ corresponds to some qubit $q$ in some set $Q_i$, as well as some height along the interval used to thicken.  The cells in its boundary comprise three possibilities.  First, the $1$-cell in $\cA$ corresponding to that qubit $q$ at the given height.  Second, any $1$-cells in $\cB_i$ corresponding to $X$-stabilizers in the boundary of that qubit $q$ at the corresponding height, and there are at most $q_X$ such $X$-stabilizers.  
Third, if a disc is attached at the given height, possibly some $1$-cells in the interior of that disc, but there are only $O(1)$ such $1$-cells.
The direct $Z$-stabilizers have weight bounded by $w'_Z$.

\item[4.] By construction of the cellulation of the disc, every edge in the cellulation has one $O(1)$-vertices in its boundary.
Every qubit in the cone code corresponding to some $1$-cell of $\cA$ is in its at most $q_Z$ stabilizers of the cone code.

\item[5.] All $X$-stabilizers correspond either to $X$-stabilizers of $\cA$, which have weight bounded by $w_X$, or $-1$-cells of $\hat \cB_i$, which are constructed to have weight $O(1)$.

\item[6.] By construction, same as item 5.

\item[7.] Consider the celled code.
Any $1$-chain $u$ in $\Cn(f)$ is a sum of a $1$-chain $q$ in $\cA$ and some $0$-chains in $\hat\cX_i$.
This $1$-chain $q$ has an obvious projection $q_i$ onto $\cQ_i$.
For each $i$, for the coboundary of $u$ to vanish, there must be some chain $v$ in $\cX_i$ such that $\partial^T v$ is equal to $q_i$.  
To see this, note
 that there is some chain $v$ in $\hat \cX_i$ such that $\partial^T v=q_i$.
 If the coboundary of $u$ vanishes on $\cQ_i^\perp$,
 we can add a coboundary of a chain in $\cR_i$ to $v$ so that the result is supported in $\cX_i$.
To see this, 
regard $v$ as an {\it edge chain}\footnote{An edge chain is our term for a $1$-chain on $\cS$.}.
This edge chain $v$ may have some support on edges of graph $G$ in the interior of the attached disc which themselves attach to vertices in $G_i$.
However, if the chain is closed in the interior of the discs, we can add boundaries of faces of $G_i$ to remove its support on those edges, pushing the chain off the discs.
Then, for the coboundary of $u$ to vanish on $\cQ_i$, indeed there must be some chain $v$ in $\cX_i$ such that $\partial^T v$ is equal to $q_i$.

Hence, $q$ must correspond also to a coclosed chain in the chain complex corresponding to the induced code, i.e., an $X$-logical operator of the induced code.  Conversely, if $q$ is a coclosed chain in the chain complex of the induced code, then such a chain $v$ does exist.  Further, such a chain $v$ is unique up to adding coboundaries of a chain in $\cR_i$ because each $\overline \cB_i$ has trivial zeroth homology.

By adding coboundaries of elements of $\cA_0$, we can shift $q$ by $X$-stabilizers of the induced code.
This shows $\tilde K=K$: given $q$, $u$ is unique up to coboundaries, by adding coboundaries in the cone code, $q$ can be shifted by coboundaries in the chain complex of the induced code, and $q$ must be coclosed in the chain complex of the induced code.

Further, if $u$ is nontrivial, then $q$ is nontrivial, and so in that case $|u|\geq d_X$.

\item[8.]  Consider the celled code.  Let $u$ be any  closed $1$-chain in $\Cn(f)$, $u=u^q+\sum_i u^x$, where $u^q$ is in $\cA_1$ and  $u^x_i\in \overline\cB_i[-1]_0$.
By assumption that $\overline \cB_i$ has trivial zeroth homology, we can add a boundary of some $v^q_i\in \cB_i[-1]_{1}$ to $u$, giving some chain $u'$ whose projection onto
$\cB_i[-1]_0$ vanishes.
We can choose further $|v^q_i|\leq |V_i|/2$, as if not, simply add the sum of all cells to $v^q_i$.
By assumption on soundness, $|u'|\leq |u|/\lambda$.

So, $u'$ must correspond to a closed chain of the induced code.
Further, by adding a boundary of $v^q_i$, where $v^q_i$ is the sum of all $2$ cells in $\cB_i[-1]_{1}$, we can shift $u'$ by a stabilizer of the induced code.
Hence, if $|u'|\leq d_Z$, we can add stabilizers to make $u'=0$.
Hence, if $u$ is a nontrivial logical, $|u|\geq d_Z \lambda$.
\end{itemize}
\end{proof}
\end{lemma}

\subsection{Examples With Toric Code and Geometric Interpretation}
\label{geometric}
Let us give a simple example of coning.  
We emphasize to the reader that this example may be {\it too simple}.  It shows how coning works, and it shows how in this particular example coning corresponds to a familiar construction with the toric code; however, the reader should not be misled into thinking that in general coning has such a simple geometric interpretation.

\begin{figure}
\includegraphics[width=3in]{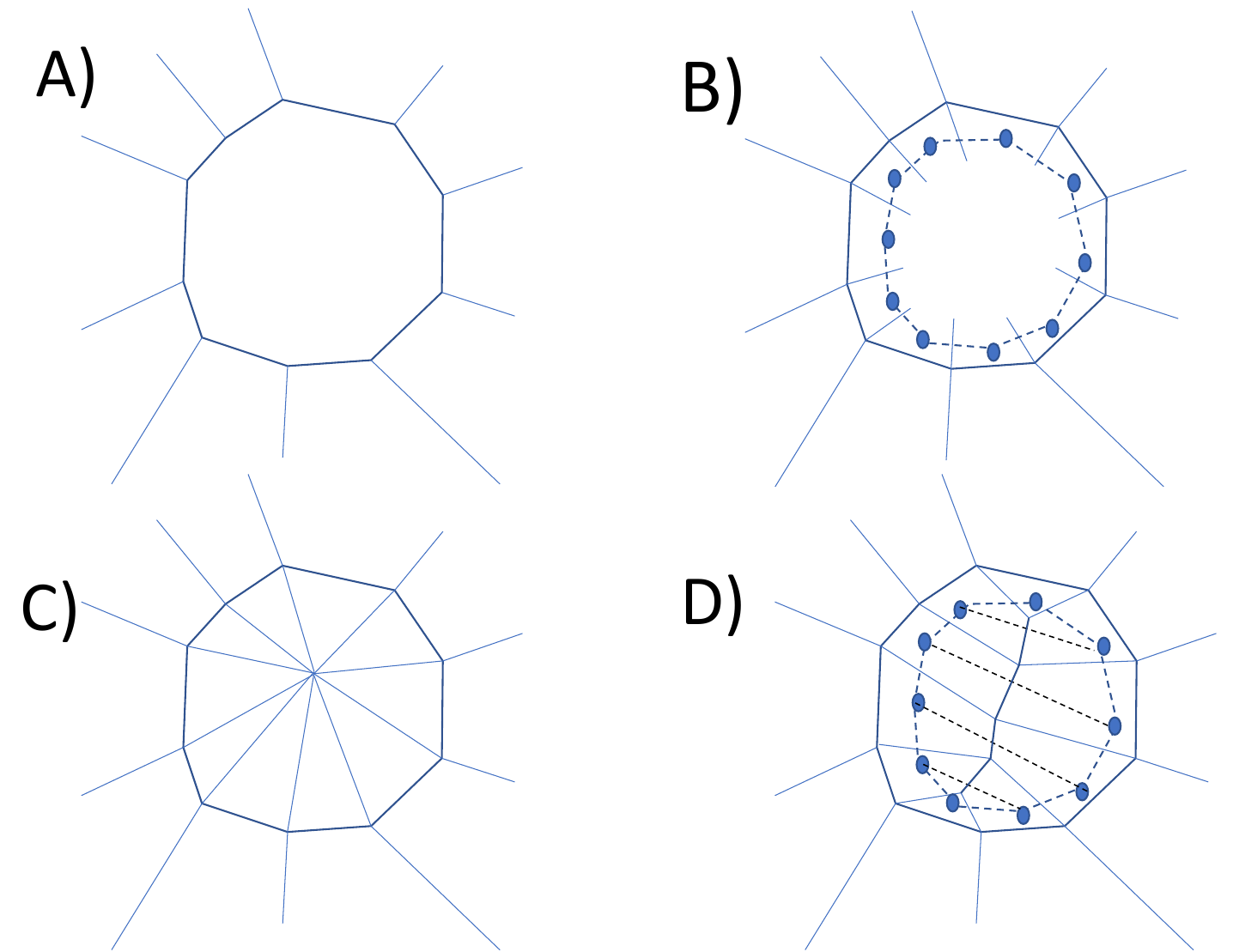}
\caption{{\bf A}: part of a cellulation of a toric code.  We show a polygon representing a high weight $Z$-stabilizer.
{\bf B}: The result with ``rough" boundary conditions if we use $\cB_1$ rather than $\overline \cB_1$ in the cone construction.
Solid lines represent qubits, vertices represent $X$-stabilizer.  There are $Z$-stabilizers on triples of lines inside the polygon, corresponding to two neighboring edges entering the polygon and one edge in the polygon between those edges.
The dashed line shows the graph $G_1$.  The edges of this graph correspond to $0$-cells of $\cB_1$ and the vertices (shown as solid circles) correspond to $1$-cells.  {\bf C:} Cone code with one additional high weight $X$-stabilizer.  Note the relation with a dual cellulation of B: the dashed line of B bounds the attached $2$-cell in $\cS_1$, which dually becomes a $0$-cell in C.
{\bf D}: Reduced cone code.  Dashed lines show the cellulation of complex $\cS_1$.  The resulting stabilizers in the reduced cone code are on the dual cellulation.}
\label{figexample}
\end{figure}

Consider a toric code, i.e., a code defined by some cellulation of a surface, with $X$-stabilizers, qubits, and $Z$-stabilizers associated with $0$-, $1$-, and $2$-cells, respectively.
Suppose we choose, however, a strange cellulation of the surface in which one of the faces is a polygon with $n$-sides for some very large $n$.
We wish to induce this stabilizer, with other $Z$-stabilizers being direct $Z$-stabilizers. 
Label qubits corresponding to edges on the boundary of this face by integers $j\in\{0,\ldots,n-1\}$.
There are $n$ $X$-stabilizers of the form $X_{j} X_{j+1} \ldots$, where the index $j$ is taken to be periodic in $n$ so that $X_n=X_0$, and the dots $\ldots$ indicates other qubits in the support of the stabilizer, other than those which correspond to
edges of this face.
We wish to induce the $Z$-stabilizer $Z_0 Z_1 \ldots Z_{n-1}$.
See Fig.~\ref{figexample}, A.

Let $Q_1$ be the set of qubits correspoding to edges of this polygon.
The complex $\cB_1$ has nontrivial first homology group.  Indeed, the sum of all $1$-cells has vanishing boundary.
So, to follow the construction above, we need to construct $\overline \cB_1$.  Suppose, however, that we did not do this, and used $\cB_1$ rather than $\overline \cB_1$ to construct the mapping cone.  Let us see what goes wrong.

We will find that we have now an additional $n$ qubits, one for each vertex on the polygon.  Label these added qubits by $\overline j$ for $j\in \{0,\ldots,n-1\}$.  Then, the stabilizers $X_j X_{j+1} \ldots$ are replaced by
$X_j X_{j+1} X_{\overline j} \ldots$ in the cone code.  Additionally,
we have  $Z$-stabilizers $Z_{\overline j} Z_{j+1} Z_{\overline{j+1}}$ for each $j$.

This is a familiar result: it is a puncture in a toric code with so-called ``rough" boundary conditions around the puncture.
This may increase the degeneracy of the ground state: if we create more than one such puncture, it will increase the degeneracy, but a single such puncture does not.
See Fig.~\ref{figexample}, B.

So, indeed we must construct $\overline \cB_1$.  In this case, we have a single redundancy in $X$-stabilizers of $\cB_1$.  Adding this redundancy, we find that the cone code adds one more $X$-stabilizer: $\prod_j X_{\overline j}$.
See Fig.~\ref{figexample}, C.

The result is again a toric code, but now with a large $X$-stabilizer weight.
This weight can be reduced in weight by cellulating the added disc in $\cS_1$.
It is a nice exercise to see that (because of the interchange $j \leftrightarrow (2-j)$ of dimension of cells between $\cS_i$ and the cone code) the cellulation of a disc in $\cS_1$ means that the reduced cone code is a toric code with the dual cellulation (in the of a dual graph) used to resolve the high weight $X$-stabilizer.
See Fig.~\ref{figexample}, D.

This example also shows the importance of the soundness factor $\lambda$.  In this case, the soundness factor $\lambda$ is of order $1/n$.  To give an example where this factor correctly estimates the effect on distance, consider a toric code given by a cellulation of an $L$-by-$L$ torus (so that there are $K=2$) logical qubits, using the usual cellulation with square $2$-cells.  Remove an $m$-by-$m$ patch of squares (including removing the edges inside the patch), and replace that patch with a single large face, i.e., adding the product of Pauli $Z$ operators on edges around that patch as a $Z$-stabilizer.  This is our desired induced code.  Then, this is precisely the situation considered above, with $m \propto n$.  The toric code with square cells has distance $L$.  Removing that patch and replacing it with a single face can reduce the distance $d_X$ because a coclosed $1$-chain can take a ``shortcut" going directly across the patch; indeed, if we had many such patches, we could potentially reduce the distance by a factor proportional to $m$.  However the distance $d_Z$ is not reduced because a closed $1$-chain must ``go around" the patch without any shortcut.  However, in the reduced cone code, the distance $d_Z$ may also be reduced since closed $1$-chains can now also take a ``shortcut" across the patch, and if we had many such patches, the reduction in distance could also be proportional to $m$, i.e., the distance may be multiplied by something proportional to $\lambda$.

\subsection{Unreasonable Codes}
\label{unreasonable}
We now consider the case of unreasonable codes.  We call the procedure ``connecting".

For simplicity, let us assume that the code has the property that for any $Z$-stabilizer $S_Z$ and any $X$-stabilizer $S_X$, we have $|\supp(S_Z) \cap \supp(S_Z)|$ equal to $0$ or $2$.  Recall that we can always achieve this by weight reducing the $X$-stabilizers.  Further, our procedure will work for many codes which do not have this property.

Suppose we have some code $C$ that includes a stabilizer $S_i$ which 
 is the product of Pauli $Z$-operators on some set of qubits $Q_i$
such that the graph $G_{Q_i}$ has two connected components (the generalization to more connected components will be obvious, and will be given below), so that the procedure above does not work.

For notational simplicity, first consider the case of a single such stabilizer.

Define the ``connected code" to be the following code: it has a new qubit, called $r$.
Pick one qubit from each of the two connected components of $G_{Q_i}$.  Call these qubits $q_1,q_2$.
We modify the $X$-stabilizers, by multiplying every $X$-stabilizer of $C$ that has support on $q_1$ or $q_2$ by $X_r$, i.e., the boundary of the cell corresponding to $r$ is the sum of boundaries of the cells corresponding to $q_1$ and $q_2$.  The $X$-stabilizers with no support on $q_1$ or $q_2$ are unchanged.
Let the connected code have the same $Z$-stabilizers as $C$, except add the stabilizer
$Z_{q_1} Z_{q_2} Z_r$ and replace $S$ with $S Z_{q_1} Z_{q_2} Z_r$.

Now, on the connected code,
 rather than inducing the stabilizer on set $Q_i$, induce a stabilizer on $(Q_i\setminus\{q_1,q_2\}) \cup \{r\}$, i.e., induce $S Z_{q_1} Z_{q_2} Z_r$.
 Let this set $(Q_i\setminus\{q_1,q_2\}) \cup \{r\}$ be denoted $Q'_i$.
For the connected code, the graph $G_{Q'_i}$ is connected and further
still has the property that the cardinality of intersection of $Z$ and $X$ stabilizers is $0$ or $2$ , so coning works.

If $G_{Q_i}$ has $k>2$ connected components, the generalization is to define the connected code by picking several pairs of distinct qubits $q'_1,q_2$ and $q'_2,q_3$ and so on, up to $q'_{k-1} q_k$, with $q_j,q'_j$ in the $j$-th connected component.  Note that we have written the first pair as $q'_1,q_2$, rather than $q_1,q_2$.  
Then, add qubits $r_1,r_2,\ldots,r_{k-1}$ and stabilizers $Z_{q'_a} Z_{q_{a+1}} Z_{r_a}$ for each $a\in \{1,\ldots,k-1\}$.
Every $X$-stabilizer with support on $q'_a$ or $q_{a+1}$ will act also on $r_a$.
Finally, induce a $Z$-stabilizer on $Q_i \setminus \{q'_1,q'_2,\ldots,q_2,q_3,\ldots\} \cup \{r_1,r_2,\ldots\}$.
Again, the graph corresponding to this stabilizer for the connected code is now connected.

If there are several $Z$-stabilizers $S_i$ with corresponding graphs $G_{Q_i}$ which have more than one connected component, we do this same procedure for each such stabilizer to define the connected code.  For each such stabilizer $S_i$, we add $k-1$ new qubits to define the connected code, corresponding to the qubits called $r$ or $r_1,r_2,\ldots$ above.  Call these added qubits the ``connecting qubits".

Let us explain why this works.  One can see that the number of logical qubits is unchanged; we have increased the number of physical qubits by one but also increased the rank of the stabilizer group by one.
Further, given any $1$-chain or $1$-cochain of the connected code, we can add a boundary or coboundary to make it vanish on all connecting qubits; indeed, for each connecting qubit there is an $X$-type stabilizer and a $Z$-stabilizer, both of which have support on that connecting qubit (and no other connecting qubits).  
Thus, given any logical operator $O$ of the connected code, we can multiply it by a stabilizer to get a logical operator $O'$ which is supported on the qubits other than the connecting qubits.
We claim that if $O$ is a nontrivial logical of the connected code, then $O'$ must be a nontrivial logical operator of the original code $C$: if $O'$ is a trivial logical of $C$, $O'$ is a product of stabilizers of $C$, and so we can multiply $O$ by the corresponding stabilizers of the connected code to get a logical operator supported just on the connecting qubits, which must then be the identity operator if it is to commute with the stabilizers.

Now consider the effect on distance.

Consider any $Z$-logical operator of the connected code.  If it has support on $r$, we can multiply by stabilizer $Z_{q_1} Z_{q_2} Z_r$ to make it not have support on $r$.
Similarly, for any $X$-logical operator, if it has support on $r$, we can multiply by any $X$-stabilizer acting on $q_1$ (or $q_2$), so that the result does not have support on $r$.
This multiplication can change the weight, but it increases the weight of a $Z$-logical operator by at most an $O(1)$ multiplicative factor and increases the weight of an $X$-logical operator by an at most $w_X$ multiplicative factor.
So, $d_X$ is reduced by an at most $O(w_X)$ factor and $d_Z$ is reduced by an at most $O(1)$ factor.

So, we may induce stabilizers even in unreasonable codes.  Indeed:
\begin{lemma}
\label{conncode}
The connected code has the same number of logical qubits as $C$.  The number of physical qubits of the connected code is $O(N)$.
The $X$- and $Z$-distances of the connected code are $\Omega(d_X),\Omega(d_Z)$ respectively.
\end{lemma}

To give some intuition for this connecting procedure, consider a toric code again.  Suppose the toric code is on a plane with several punctures with some boundary conditions on each puncture.  Suppose boundary conditions for the punctures are chosen in some way so that the product of Pauli $Z$ around each puncture is a logical operator.  For $k\geq 2$ punctures, called $1,2,\ldots,k$, let $L_1,L_2,\ldots,L_k$ be the corresponding products of Pauli $Z$-operators.  Suppose further that we wish to induce the product $L_1 L_2 \ldots L_k$ as a $Z$-stabilizer, without separately inducing any other stabilizers.

One natural way to do this is to attach a $k$-punctured sphere to those $k$-punctures.  This doesn't quite work, but let's delve into this further.
Suppose $k=2$, so we want to attach a twice punctured sphere.  We can imagine doing this in two steps: first use a ``pair of pants" to turn the two punctures into a single puncture, then close the pair of pants with a disc.  One 
simple version of a pair of pants is the following: let $e_1$ be an edge in one puncture, and $e_2$ be an edge in the other.  Add two new edges $f,g$ so that $e_1,e_2,f,g$ make a plaquette (this means that the edges $f,g$ are not drawn in the plane!).  This plaquettes means that we add a $Z$-stabilizer $Z_{e_1} Z_{e_2} Z_f Z_g$.  There is now a single closed loop obtained by taking the two boundaries of the punctures, cutting out edges $e_1,e_2$ from the punctures, and adding in edges $f,g$ and we can attach 
a disk to this loop.  Unfortunately, while this pair of pants operation turns the two punctures into a single puncture, the result adds unwanted first homology.  Consider for example the toric code on a sphere, with two punctures, each puncture consisting of a single plaquette.  This code has a single logical qubit (we have removed two $Z$-stabilizers from the sphere, but one of them is redundant).  Adding in the product of Pauli $Z$-operators around both punctures does {\it not} change the stabilizer group in this case, so there should still be a single logical qubit.  However, the above pair of pants procedure instead produces a torus which has {\it two} logical qubits.  The trouble is indeed this added extra loop.
In \cite{fh}, a similar issue was considered (see discussion in section 1.3); the resolution there was roughly to consider higher dimensional toric codes.  Here our resolution is even simpler, replacing $f,g$ with a single qubit.  This can be understood (very heuristically) in two different ways.  One is to think of this pair of pants construction but restrict to the subspace with $X_f=X_g$ which removes this unwanted first homology.  The other (even more heuristically, and perhaps this will not be useful to the reader) is to think that it corresponds to considering higher dimensional codes similar to the construction in \cite{fh}.

Now let us discuss the application of connecting to implementing logical Clifford operations which are composed of elementary logical CNOTs (i.e., logical Clifford which map Pauli $Z$ operators to products of Pauli $Z$) on some arbitrary LDPC CSS stabilizer code $C$.  First, as well-known, given two copies of some code, the CNOT gate can be implemented transversally.  Suppose we have three or more copies of a code, labelled $1,2,3,\ldots$.  If we start with a maximally entangled state between copies 2 and 3, given by preparing the logical qubits of 2,3 in Bell pairs and then applying an arbitrary Clifford $U$ to 3, we can then apply a transversal CNOT from 1 to 2, measure all qubits of 1 in the $X$ basis (hence measuring all the logical $X$ operators of $X$ in a fault tolerant way) and measure all qubits of 2 in the $Z$ basis to teleport the state from 1 to 3, while applying the Clifford $U$.  The teleportation is up to some Pauli correction on the logical qubits of 3, which can be implemented by a product of Paulis on the physical bits.  This reduces the task of implementing logical Cliffords to the task of preparing appropriate entangled states between 2 and 3.  However, such an entangled state, for the given Cliffords, is the unique code state of a code $C_{ent}$ with no logical qubits: this code $C_{ent}$ is given by taking two copies of the given stabilizer code and adding some additional stabilizers which are products of logical operators between the copies; these additional stabilizers include both $X$- and $Z$-type stabilizers and so we need to induce both such.  We can  use the combining procedure to implement these additional stabilizers by constructing some reduced cone code for $C_{ent}$.  Using the methods of \cref{impsound}, we will see that we can assume that the soundness of $C_{ent}$ is $\Omega(1)$.
Remark: it would be interesting also to induce stabilizers which are products of Pauli $X$ in one copy with Pauli $Z$ in another to allow performing additional logical operations; we suspect that this is possible but leave it open.

This leaves a few details to consider, which we only sketch here.  
First, the added stabilizers to define the code $C_{ent}$ are not only high weight (which we deal with using combining and coning as explained here) but may also lead to a large $q_Z$.  Indeed, $q_Z$ may potentially be as large as the number of logical qubits of the code.  So, it may be necessary to thicken the code throughout to reduce $q_Z$ to $O(1)$.

Second, what we prepare is not the code state of $C_{ent}$, but rather of some reduced cone code.  This reduced cone code is given by some further thickening of two copies of $C$, with some added qubits and added stabilizers, and also some modification of the $X$-stabilizers of the thickened copies so that they involve also the added qubits.
Let us call the qubits of the reduced cone code either ``original" or ``added".
We can then measure all the added qubits in the $X$-basis, so that then the remaining original qubits are stabilized by the thickened $C_{ent}$, up to some Pauli correction, depending on the outcome of the measurement.  So, this measurement produces the desired entangled state on two thickened copies of $C$, so we should work with thickened copies of $C$ throughout, rather than $C$.  From here on we will simply refer to $C$ and $C_{ent}$ for brevity, rather than constantly repeating that the codes are thickened.

A third detail is that one might also like to be able to measure some set of the logical qubits of some code, while leaving the others unaffected.  We can do this by a generalization of the above procedure.  Use four copies, $1,2,3,4$, where 1 encodes some logical state of $C$ and $2,3,4$ are copies of $C$ with some of the logical qubits of 2 in Bell pairs
with the corresponding logical qubits of 3 and the other logical qubits of 2 in Bell pairs with the corresponding logical qubits of 4 (the logical qubits of 3,4 which are not in Bell pairs with logical qubits of 2 may be taken to be in the logical $Z=+1$ state).  Then, again CNOT from 1 to 2, and measure 1 and 2 as before, thus teleporting some qubits of 1 to 3 and the other qubits of 1 to 4.  Finally, one may measure are physical qubits of 4 in the $Z$-basis (or $X$-basis) to measure some subset of the logical qubits of the initial state of 1 in the $Z$-basis (or $X$-basis) while leaving the others intact by teleporting them to 3.

The most important question is whether this procedure is fault tolerant.  One model to analyze fault tolerance in is some kind of circuit model where qubits are subject to noise while stabilizers are repeatedly measured, possibly with noisy measurement of stabilizers.  Of course, we do not have even an analysis of fault tolerance for the code $C$ itself, much less $C_{ent}$: all we know is that $C$ is a large distance code.  
So we may consider a simpler question: suppose we prepare the reduced cone code assuming perfect syndrome measurement, and then some number of errors are applied to the qubits of the reduced cone code, and next we measure the added qubits of the reduced cone code in the $X$-basis. 
We finally measure the stabilizers of the two copies of $C$ (we cannot measure the additional stabilizers of $C_{ent}$ because those are high weight). Errors applied to original qubits of the reduced cone code correspond simply to errors on the two copies of $C$, and so these errors can be corrected after the final stabilizer measurement if the number of them is small enough compared to distance.  $X$-errors on the added qubits of the reduced cone code have no effect since we measure those qubits in the $Z$-basis.  What we must consider are $Z$-errors on the added qubits of the reduced cone code.  However, up to boundaries, these $Z$-errors on the added qubits of the reduced cone code are equivalent to $Z$-errors on the original qubits, with the total weight of the error on the original qubits being at most multiplied by the inverse soundness factor.  Since we may assume that the soundness is $\Omega(1)$, this bounds the error weight on the original qubits, so those errors can also be corrected after the final stabilizer measurement if the number of errors on the added qubits was sufficiently small compared to distance.

\section{Putting It Together}
\label{together}
We claim:
\begin{theorem}
\label{putthm}
Let $C$ be any code, with some given parameters, $w_X,w_Z,q_X,q_Z$.
Let $\poly(\dev)$
refer to a quantity that is polynomial in $w_X,w_Z,q_X,q_Z$, where ``dev" is shorthand for ``deviation from LDPC".
Then there is an LDPC code, whose parameters we will denote with a tilde, such that $\tilde K=K$, and with $\tilde
 N$ bounded by $N$ times $\poly(\dev)\polylog(N)$ and with $\tilde d_X,\tilde d_Z$ lower bounded by $d_X,d_Z$ divided by 
 $\poly(\dev)\polylog(N)$.
 
 If the input code is LDPC, with arbitrary $w_Z,w_X,q_X,q_Z=O(1)$, we may choose the tilde code to have
 $w_X\leq 5$, $q_X\leq 3$, $w_Z \leq 5$, and $q_Z\leq 5$.
 \end{theorem}
 
 We do this by using copying, gauging, thickening, and coning.  We will not write tildes above parameters because we will be applying several different transformations to the code, and it would then necessitate write multiple tildes to keep track of all these steps.  Instead, we write parameters without tilde through, but we describe the parameters ``before" and ``after" various steps.
 
First, we copy and gauge to reduce $q_X,w_X$ to $O(1)$.  Then thicken to reduce $q_Z$ to $O(1)$.  
Note that since $q_X,w_X$ are both $O(1)$ after copying and gauging, after thickening all parameters $q_X,w_X,q_Z$ are all $O(1)$; it was useful to do copying and gauging first, as if we thicken a code with $w_X=\omega(1)$, then
the resulting code will not have $w_Z=O(1)$.
 
Finally, construct the reduced cone code to reduce $w_Z$ to $O(1)$.
 Stabilizers corresponding to elements of $\cC_1 \otimes \cE_1$ can be direct if desired.
 
 Let us track the parameters in detail, assuming that the input code is LDPC (but with some large parameters).  After copying and gauging we have $w_X,q_X\leq 3$.
 After thickening we have $w_X\leq 5$ and $q_X\leq 3$.  From \cref{qsplitlemma2}, if the input code is LDPC so that $q_Z w_Z=O(1)$, we may take any $w>0$ and still have $\ell=O(1)$.  Taking $w=3$ we get a code with $q_Z\leq 5$; there is no point in taking smaller $w$.
 For the reduced cone code, we may cellulate the discs with pentagons so that still $w_X\leq 5$ and $q_X\leq 3$ and also we may have $q_Z\leq 5$.
 We pick code $C'$ to have no $Z$-stabilizers.  Then, the reduced cone code may be constructed to have
 $w_Z\leq 5$; see item 3 of \cref{conecode}, where the $O(1)$ constant may be taken to be $1$.

Importantly, the soundness parameter $\lambda$ of some code, when applying coning, is in worst case $1/w_Z$, where here, $w_Z$ refers to the parameter for the code input to the coning step, which $O(\poly(\dev))$.
In many applications $\lambda$ may be larger: see the next section.

An interesting question is to what extent the operations we have described are homotopy equivalences of chain constants with bounded Lipchitz constants\cite{hho}.  Such a result could be used to show efficient decoding algorithms for these weight reduced codes.

\section{Coding Applications: Weight Reducing a Linear Distance Code}
\label{applic}
We consider the following construction.
We choose $X$-stabilizers to be random, of average weight $\Delta=\beta\log(N)$ on $N$ bits, for some constant $\beta$.
We choose there to be $N/2$ such stabilizers.  The particular choice of $N/2$ is arbitrary; we need $dN$ for some $d\in (0,1)$. 
We choose these stabilizers as follows: the boundary operator from qubits to $X$-stabilizers is an $N/2$-by-$N$ matrix, and
each entry is chosen independently at random, chosen to be $1$ with probability $\Delta/N$ and $0$ otherwise.

Remark: it is possible that we could instead start with $X$-stabilizers of weight $O(1)$, by using some construction of a
good classical LDPC code.  We choose the logarithmic weight as it simplifies some estimates.
However, starting with an LDPC classical code might improve the final distance estimates by some logarithmic factors.

We will then add $Z$-stabilizers below, and use coning to weight reduce those stabilizers.

We will pick $\beta$ so that the conclusions of various lemmas below follow.

\subsection{A linear distance code with $w_X,q_X=O(\log(N))$, but $w_Z,q_X=\Theta(N)$:}
We begin by constructing a code with these properties and then use thickening and coning to reduce $Z$-stabilizer weight.

\begin{lemma}
\label{classgood}
For sufficiently large $\beta$, with high probability this classical code has distance $\Theta(N)$.
\begin{proof}
This is the essentially same proof prop 3.5 of \cite{hho}, which as noted there is related to a
standard property of LDPC codes, that they achieve the
Gilbert-Varshamov bound as degree $\rightarrow \infty$.
We need $\beta$ large enough to ensure that, with high probability, all vertices have large degree.
\end{proof}
\end{lemma}

Regard this classical code as a $1$-complex $\cA$.
So, the homology group $H_1(\cA)$ has rank $\geq N/2$ since there are $N/2$ $X$-stabilizers.
The rank may be larger if there are linear dependencies among these stabilizers.
Let $b_j(\cA)$ denote the rank of $H_j(\cA)$.
We have
\begin{lemma}
\label{bettilemma}
For large enough $\beta$,
with high probability, $b_1(\cA)=N/2$.
More strongly,
the expectation value of $2^{b_1(\cA)-N/2}$ equals $1+O(1/N^{100})$.
\begin{proof}
The number of $0$-chains in $A$ with vanishing coboundary equals
$2^{b_0(\cA)}$.  By prop 3.4 of \cite{hho}, this number equals $1+O(1/N^{100})$; this result is not in the statement of the lemma,
but follows from the proof which counts the number of nonzero chains with nonvanishing coboundary.
Remark: \cite{hho} instead used $(3/4)N$ stabilizers rather than $N/2$; the exact fraction of stabilizers is unimportant so long
as $\beta$ is large enough.

However, $b_1(\cA)-N/2=b_0(\cA)$, by Euler characteristic.
\end{proof}
\end{lemma}

Choose $N/4$ elements of $H_1(\cA)$ independently and uniformly at random (one can choose to include the trivial element or not; it only is chosen with probability $o(2^{-N/2})$ so it does not matter if we include it or not).
The corresponding $Z$-type operators will be the $Z$-stabilizers of the induced code.
\begin{lemma}
With high probability these $Z$-stabilizers are linearly independent.
\begin{proof}
$H_1(\cA)$ has rank $k\geq (1-d)N)$.  So, we can write the $Z$-stabilizers in terms of a $k$-by-$N/4$ matrix over $F_2$.
Columns of this matrix are chosen independently and uniformly at random (including the all $0$s column
 if we include the trivial element of $H_1(\cA)$).
So, we must show that this matrix is full rank.
The $(j+1)$-st column is linearly dependent on the first $j$ columns with probability
$\leq 2^{-(k-j)}\leq  2^{-N/4}$.  By a union bound, the lemma follows.
\end{proof}
\end{lemma}

\begin{lemma}
\label{inddist}
With high probability, this code has  distances $d_X,d_Z=\Theta(N)$.
\begin{proof}
The lower bound on $d_Z$ follows immediately from lemma \ref{classgood}.  That is, every product of Pauli $Z$-operators that commutes with the $X$-stabilizers has high weight.

Lower bounding $d_X$ is slightly different because it is not the case that every product of Pauli $X$-operators that commutes
with all $Z$-stabilizers has high weight.  Indeed, any $X$-stabilizer is a logarithmic weight operator that commutes with all $Z$-stabilizers.  So, we need to use the fact that a logical $X$-operator is not a product of $X$-stabilizers, i.e., 
it corresponds to
a $1$-chain $u$ that represents a nontrivial element of $H^1(\cA)$.
We now use an easy first moment argument.  Fix $\cA$.  Fix $u$, and then choose the $Z$-stabilizers.
For each $Z$-stabilizer, we have chosen some element of $H_1(\cA)$ at random, and, since $u$ represents a nontrivial element of $H^1(\cA)$, with probability $1/2$ the chain $u$ has nonvanishing inner product with the given element of $H_1(\cA)$.
So, the probability that $u$ has vanishing coboundary is $2^{-N/4}$.
The number of elements of $H^1(\cA)$ with weight at most $w$ is bounded by the number of vectors of Hamming weight at most $w$.
So, the expected number of $u$ with weight at most $w$ is $\leq 2^{-N/4} \sum_{j=0}^w {N\choose w}$. 
For $w$ sufficiently small compared to $N$, this is exponentially smaller.
\end{proof}
\end{lemma}

\subsection{Thickening and Coning}
\label{TandC}
Thicken to reduce $q_Z$.  By \cref{Oeps}, we can, for any $\epsilon>0$, reduce $q_Z$ to $O(1)$, while increasing from $N$ qubits to $O(N^{2+\epsilon})$ qubits, keeping $d_Z=\Theta(N)$ and making $d_X=\Theta(N^{1+\epsilon})$.
We will see that this suffices at the end of the construction to get a code on $n$ qubits with distance $\Omega{n^{2/3}-\epsilon})$ for every $\epsilon>0$.  In order to ultimately get distance $\tilde \Omega(n^{2/3})$, we instead reduce $q_Z$ to $O(\log(N))$ bu taking $\ell=\Theta(N)$.

We then apply coning to reduce $w_Z$.  Surprisingly, the soundness parameters are all $\Omega(\Delta)$:
\begin{lemma}
\label{csound}
For large enough $\beta$,
with high probability, there is a choice of pairing such that the graphs $G_i$ are connected (and hence with high probability the code is connected), and with high probability there is a choice of pairing such that the graphs $G_i$ are expanders, so that $h(G_i)=\Omega(\Delta)$
\begin{proof}
We will show that, for any given $Z$-stabilizer, there is a choice of pairing such that the desired properties of the corresponding graph $G_i$ hold with probability at least $1-o(1/N)$.  Further, we will show that a random pairing suffices.  Then, by a union bound, the lemma follows.

Consider a $Z$-stabilizer.  This is chosen to be a random element of $H_1(\cA)$, i.e., we chose $\cA$ at random, and then chose an $N$ bit vector $v$ uniformly at random, such that $\partial v=0$, and then this vector was the desired element of $H_1(\cA)$.
For given $\cA$, number of $v$ with $\partial v=0$ is $2^{b_1(\cA)}$.

So, the probability distribution on pairs $v,\cA$ is as follows: the probability is nonvanishing iff $\partial v=0$, in which case the probability is $2^{-b_1(\cA)}$, multiplied by the a priori distribution on $\cA$.  Call that a priori distribution $\rho(\cA)$.

Suppose instead we had chosen a different a priori probability distribution for $\cA$.  Suppose that we had chosen the probability distribution $\rho'(\cA)$ defined to be proportional to $2^{b_1(\cA)} \rho(\cA)$.
We will show that the desired expansion occurs with high probability given this a priori distribution $\rho'(\cA)$.
This will then imply the lemma: by lemma \ref{bettilemma}, with high probability for distribution $\rho'(\cA)$ we have $b_1(\cA)=N/2$ with probability $1-O(1/N^{100})$ and so the total variation distance between $\rho,\rho'$ is $O(1/N^{100})$.

In the second step, consider the probability distribution on pairs $v,\cA$ given that $\cA$ was chosen from $\rho'(\cA)$.
We now find that the probability is nonvanishing iff $\partial v=0$, in which case the probability equal $\rho(\cA)$.
Fix some nonzero $v$.  Choose a random pairing.  This induces a random graph, with vertices corresponding to nonzero entries of $v$.  
This random graph depends on only on a submatrix of the boundary operator, the submatrix containing columns corresponding to nonzero entries of $v$.
These entries of the submatrix are chosen independently, with each entry chosen to equal $1$ with probability $\Delta/N$, and then conditioned on there being an even number of $1$s in each row.
Each row then adds a number of edges to the graph equal to one half the number of $1$s in that row, and, for given number of $1$s, the edges are added between uniformly random pairs of vertices subject to the requirement that all edges added by a given row are between distinct vertices.
However, different rows can add edges between the same vertices.
Thus, this is almost, but not quite the same as the model of an Erd\"{o}s-Renyi random graph, due to the possibility of multi-edges and due to some correlation between edges added in a given row.
Nevertheless, it is a standard argument that the resulting graph has the given edge expansion properties with high probability, indeed with probability $o(1/N)$ for large enough $\beta$.

We give this argument for completeness.
We restrict to the case that $|v|=\Theta(N)$ for simplicity, since with high probability all $v$ with $\partial v=0$ have that.
Let $w\leq v$, with $0<|w|\leq |v|/2$ define some subset of vertices of the graph.
Consider the submatrix of the boundary operator defined by columns in $v$, i.e., those in the support of $v$.
For any row, let $n^\perp$ denote the number of $1$s in columns in $v-w$ , and let $n$ denote the number of $1$s in columns in $w$.
Then, with probability $1-o(1/N)$, every row of the given submatrix has $n^\perp=\Delta \Theta(1)$.
Consider two cases, $|w|>|v|/100$ or $|w|\leq |v|/100$.  In the first case, for large enough $\beta$, with probability $1-O(1/N^{100})$ every row also has
$n=\Delta \Theta(1)$, and with probability $1-O(1/N^{100})$, there are $\Delta \Theta(1)$ pairs from a column in $w$ to a column in $v-w$.  Hence, there are $\Delta N \Theta(1)$ edges from vertices in $w$ to those in $v-w$, showing the result in this case.

If instead $|w|\leq |v|/100$, consider the following model for a random pairing: pick a vertex in $w$, pair it randomly (possibly with a vertex in $v-w$ or in $w$).  Pick the next unpaired vertex (in some arbitrary order) in $w$, pair it randomly, and continue until all vertices in $w$ are paired.  The pairing of any remaining vertices in $v-w$ does not matter but can be done in the same way.  Then, the probability that on a given step we pair a vertex in $w$ with another vertex in $w$
is strictly smaller than $1$; in our case it is at most $1/99$.
Imagine a counter that starts at $0$ and is incremented by $2$ every time we pair vertices in $w$, so that it counts the number of
vertices in $w$ paired with a vertex in $v$
If we consider instead a model in which for each vertex in $w$ with a $1$ in the given row, the counter is incremented by $2$ with probability $1/99$, the probability that the counter is incremented is at least as large whenever we consider a previously unpaired vertex, plus in this second model the counter may be incremented when considering vertices which have already been paired.
That is, in the first model if the counter has some given probability of reaching some given count $c$, in the second model, the
probability of reaching that count is at least as large.
This second model makes the counter easier to estimate as events become uncorrelated.
However, in this model, with probability $O(1/N^{100})$, the total number of pairings of vertices in $w$, summed over all rows, is
at most $\Delta |w|$ times a constant much smaller than $1$.  On the other hand, with probability $O(1/N^{100})$, the total number of $1$s over all rows, in a column in $w$, is at least $0.99 \Delta |w|$.
So, there must be at least $\Theta(1) \Delta |w|$ pairings between a vertex in $w$ and one not in $w$.
\end{proof}
\end{lemma}

So, we have
\begin{theorem}
\label{applictheorem}
There is a family of LDPC quantum codes on $N$ qubits with distances $d_X,d_Z=\tilde \Omega(N^{2/3})$ and $\tilde \Theta(N^{2/3})$ logical qubits.
\begin{proof}
Let's first go through the construction in the case that we had thickened (see start of \cref{TandC}) by $\ell=\Theta(N^{1+\epsilon})$.
Then we will see that we can slightly improve this by thickening only by $\ell=\Theta(N)$.

Thickening by $\ell=\Theta(N^{1+\epsilon})$,
the reduced cone code has $\dev=O(\log(N))$.  It has $N=\tilde O(N^{2+\epsilon})$; note that
$\sum_i |Q_i| q_X$ is $\tilde O(N^{2+\epsilon})$.  The reduced cone code
has distance $d_X=\tilde \Omega(N^{2+\epsilon})$
and $d_Z=\tilde\Omega(N)$ and $K=\Theta(N)$.  The increase in distance $d_X$ comes from the thickening by $\ell$; see lemma \ref{prodC}.
The reduced cone code can be reduced to an LDPC code by \cref{putthm} at the cost of a polylogarithmic increase in $N$ and polylogarithmic reduction in $d_X,d_Z$. 
Apply distance balancing to obtain a code on $\tilde O(N^{3+\epsilon})$ qubits with both distances $d_X,d_Z=\tilde \Omega(N^{2+\epsilon})$.
The original distance balancing of \cite{owr} does not increase the number of logical qubits; the distance balancing
of \cite{EKZ20} improves the rate, giving $\Theta(N^{2+\epsilon})$ logical qubits.

So, this gives a distance scaling as the $2/3-\epsilon$ power of the number of physical qubits, for any $\epsilon>0$.

In order to slightly improve this, to a distance scaling as the $2/3$ power of the number of physical qubits up to polylogs, let's instead thicken by $\ell=\Theta(N)$.
Then, again the reduced cone code has $\dev=O(\log(N))$ but $N=\tilde O(N^2)$ and $d_X=\tilde \Omega(N^2)$ and $d_Z=\tilde\Omega(N)$.
Reduce to an LDPC code by \cref{putthm}, and apply distance balancing to get finally distance $d_X,d_Z=\tilde\Omega(N^{2/3})$.
The reason that the reduced thickening helps here is that when we thickened the first time, we had $w_Z=\Theta(N)$.  However, after constructing the reduced cone code, the thickenings used in \cref{putthm} had only $w_Z=O(\log(N))$.
\end{proof}
\end{theorem}

\section{Improving Soundness}
\label{impsound}
Much of the work in the above section was devoted to showing that the soundness factors $h(G)$ were not small.  That relied on a random construction of the code.  In this section, we describe a procedure to {\it increase} the soundness factor to $\Omega(1)$, at the cost of at most a constant factor increase in number of physical qubits and constant factor reduction in distance.
If the code is not reasonable, one should first apply the construction of \cref{unreasonable} to define the connected code, and then apply this construction here to the connected code.

We call the procedure here ``improving soundness".
Consider a reasonable code $C$.  For each $i$, construct $\cB_i$ as before in coning.  Define  graph $G_i$ as before.  Then, define a new complex $\cB'_i$ and new graph $G'_i$ as follows.  The graph $G'_i$
is defined by adding edges to $G_i$; the graphs $G'_i$ and $G_i$ have the same vertex set and every edge of $G_i$ is an edge of $G'_i$.  Edges are added only between vertices in the same connected component.  We add these edges so that $h(G'_i)=\Omega(1)$.  By standard constructions of expanders, this can be done by increasing the degree of each vertex by $O(1)$.

The complex $\cB'_i$ has the same set of $1$-cells as does $\cB_i$, but it has some added $0$-cells, corresponding to the added edges.
The complex $\cB'_i$ is defined from $G'_i$ in the obvious way.  $0$-cells of $\cB'_i$ correspond to edges of $G'_i$ and $1$-cells correspond to vertices of $G'_i$, with a $0$-cell in the boundary of a $1$-cell if the corresponding edge has the corresponding vertex in its boundary.
Define a chain map $f'_i:\cB'_i\rightarrow \cA$ so that $f'_i$ vanishes on the added $0$-cells and otherwise agrees with $f_i$.

Then, proceed as before, with complexes replaced with their primed version:
construct $\overline \cB'_i$ from $\cB'_i$, construct $\overline f'_i$ from $f_i$, and so on.

In this way, we obtain a reduced cone code with the same properties as \cref{conecode}, except the soundness factor is $\Omega(1)$.

Remark: this procedure allows a simplification of the construction of the previous section, as there is no need to worry about bounding the soundness of the various graphs as in \cref{csound}.  However, I did not discover this improving soundness construction until after doing the construction of the previous section, so I leave it as is.

Remark: further, this procedure allows a partial derandomization of the construction of the previous section and allows an improvement in the polylogs in the distance, as the code defining $\cA$ can be chosen to be any good classical LDPC code.

Remark: this procedure adds some number of extra qubits to the reduced cone code, corresponding to edges added to each $G'_i$.  However, this procedure adds also $X$-stabilizers, as the complex $\cB'_i$ will have a larger zeroth homology group than $\cB_i$ does.

{\it Acknowledgments---} I thank M. Freedman, J. Haah, and G. Z\'emor for useful discussions.  I thank A. Wills for useful comments and corrections.
\bibliographystyle{alpha}
\bibliography{qwr-ref}
\end{document}